\documentclass[11pt]{article}

\usepackage[left=2cm, right=2cm, top=2.5cm, bottom=2.5cm]{geometry}
\usepackage[x11names]{xcolor}
\usepackage{fancyhdr, amssymb, cancel, amsmath, graphicx, pgfplots, tikz}

\newcommand{\stylecolor}{black}

\usepackage[colorlinks=true, urlcolor=violet, linkcolor=blue, citecolor=red, hyperindex=true, linktocpage=true]{hyperref}

\usepackage[explicit]{titlesec}

\newcommand*\sectionlabel{}
\titleformat{\section}
  {\gdef\sectionlabel{}
   \Large\bfseries\scshape}
  {\gdef\sectionlabel{\thesection. }}{0pt}
  {\begin{tikzpicture}[remember picture,overlay]
	\draw (-0.2, 0) node[right] {\textsf{\sectionlabel#1}};
	\draw[thick] (0, -0.4) -- (\textwidth, -0.4);
       \end{tikzpicture}
  }
\titlespacing*{\section}{0pt}{15pt}{20pt}

\newcommand*\subsectionlabel{}
\titleformat{\subsection}
  {\gdef\subsectionlabel{}
   \large\bfseries\scshape}
  {\gdef\subsectionlabel{\thesubsection.\ \  }}{0pt}
  {\begin{tikzpicture}[remember picture,overlay]
    	\draw (-0.15, 0) node[right] {\textsf{\subsectionlabel#1}};
       \end{tikzpicture}
  }
\titlespacing*{\subsection}{0pt}{10pt}{10pt}

\newcommand*\subsubsectionlabel{}
\titleformat{\subsubsection}
  {\gdef\subsubsectionlabel{}
   \bfseries\scshape}
  {\gdef\subsubsectionlabel{\thesubsubsection.\ \  }}{0pt}
  {\begin{tikzpicture}[remember picture,overlay]
    	\draw (-0.15, 0) node[right] {\textsf{\subsubsectionlabel#1}};
       \end{tikzpicture}
  }
\titlespacing*{\subsubsection}{0pt}{7pt}{7pt}

\pgfplotsset{every axis legend/.append style={at={(1.02,1)},anchor=north west}}

\newcommand{\titletext}{Conductivity of weakly disordered strange metals: \\ from conformal to hyperscaling-violating regimes}

\begin{document}

\allowdisplaybreaks

\pagestyle{fancy}
\renewcommand{\headrulewidth}{0pt}
\fancyhead{}

\fancyfoot{}
\fancyfoot[C] {\textsf{\textbf{\thepage}}}

\begin{equation*}
\begin{tikzpicture}
\draw (0.5\textwidth, -3) node[text width = \textwidth] {{\huge \begin{center} \color{\stylecolor} \textsf{\textbf{\titletext}} \end{center}}};
\end{tikzpicture}
\end{equation*}
\begin{equation*}
\begin{tikzpicture}
\draw (0.5\textwidth, 0.1) node[text width=\textwidth] {\large \color{black} $\text{\textsf{Andrew Lucas$^a$ and Subir Sachdev$^{a,b}$}}$};
\draw (0.5\textwidth, -0.5) node[text width=\textwidth] {\small  $^a$\textsf{Department of Physics, Harvard University, Cambridge, MA 02138, USA}};
\draw (0.5\textwidth, -1) node[text width=\textwidth] {\small  $^b$\textsf{Perimeter Institute for Theoretical Physics, Waterloo, Ontario N2L 2Y5, Canada}};
\end{tikzpicture}
\end{equation*}
\begin{equation*}
\begin{tikzpicture}
\draw (0.5\textwidth, -6) node[below, text width=0.8\textwidth] {\small We present a semi-analytic method for constructing holographic black holes that interpolate from anti-de Sitter space to hyperscaling-violating geometries.   
These are holographic duals of conformal field theories in the presence of an applied chemical potential, $\mu$, at a non-zero temperature, $T$, and allow us to describe the crossover from `strange metal' physics at $T \ll \mu$, to conformal physics
at $T \gg \mu$. Our holographic technique adds an extra gauge field and exploits structure of the Einstein-Maxwell system to manifestly find 1-parameter families of solutions of the Einstein-matter system in terms of a small family of functions, obeying a nested set of differential equations.   Using these interpolating geometries, we re-consider holographically some recent questions of interest about hyperscaling-violating field theories.  Our focus is a more detailed holographic computation of the conductivity of strange metals, weakly perturbed by disorder coupled to scalar operators, including both the average conductivity as well as sample-to-sample fluctuations.  Our findings are consistent with previous scaling arguments, though we point out logarithmic corrections in some special (holographic) cases.   We also discuss the nature of superconducting instabilities in hyperscaling-violating geometries with appropriate choices of scalar couplings.};
\end{tikzpicture}
\end{equation*}
\begin{equation*}
\begin{tikzpicture}
\draw (0, -13.1) node[right, text width=0.5\paperwidth] {\texttt{lucas@fas.harvard.edu \\ sachdev@g.harvard.edu}};
\draw (\textwidth, -13.1) node[left] {\textsf{\today}};
\end{tikzpicture}
\end{equation*}

\tableofcontents

\section{Introduction}

A powerful way of realizing a non-Fermi liquid `strange metal' is by applying a chemical potential to a strongly-coupled
conformal field theory (CFT). In a holographic setting, the infra-red (IR) geometry is then represented by a charged black 
brane. If we restrict attention to systems with power-law correlations in the IR (as is expected for strange metals), then 
the most general isotropic IR geometry at zero temperature ($T$) is of the `hyperscaling violating' variety \cite{charmousis,iizuka,ogawa,lsb}.
This geometry captures the key characteristics of non-Fermi liquids including the finite compressibility, and an entropy density
which vanishes as a power of $T$ (Friedel oscillations representing the 
presence of an underlying Fermi surface are however not present in the classical gravitational theory, although there are arguments 
that they should appear upon including quantum gravity corrections \cite{nabil1,ssmonopole,gopakumar,nabil2}).

With a holographic representation of the non-Fermi liquid in hand, it becomes possible to address transport properties in a framework
which makes no reference to quasiparticle excitations. However, in a configuration with full translational symmetry, the conservation
of momentum, and the non-zero cross-correlation between momentum and electrical current in the presence of a chemical potential
together imply that the electrical resistivity $\rho_{\mathrm{dc}}$ is zero; so we must break translational invariance to obtain
a non-zero $\rho_{\mathrm{dc}}$ \cite{hkms}. A variety of routes to breaking translational symmetry have been 
investigated in the literature, including random potentials \cite{hkms, herzog, adams1, adams2, hartnollhofman, arean, Lucas},
periodic potentials \cite{hartnollhofman, santos1, schalm, santos2, santos3, cls, ling, donos1409}, massive gravity \cite{vegh, davison, blake1, dsz} and Q-lattices which explicitly preserve homogeneity while breaking translational symmetry (and other related methods) \cite{donos1, andrade, donos2, gouteraux2}.   Here, we focus on the influence of a weak random potential, of mean-square variance $\varepsilon^2$, 
coupling linearly to an operator $\mathcal{O}$ in the boundary theory. Then, a solution of the bulk gravitational theory
shows that \cite{Lucas}
\begin{equation}
\rho_{\mathrm{dc}} \sim \varepsilon^2 T^{2(1+\Delta-z)/z} \label{rhomain}
\end{equation}
where $z$ is the dynamic critical exponent of the hyperscaling violating geometry, and $\Delta$ is the scaling dimension
of the operator $\mathcal{O}$ (the precise $T$ range over which Eq.~(\ref{rhomain}) holds is described below).
Remarkably, Eq.~(\ref{rhomain}) coincides with the result obtained by a ``memory function'' analysis of the boundary field theory \cite{hkms,hartnollhofman,Lucas,santos2}, without the use of gravity.

We note that these holographic advances have recently inspired a re-examination of the field-theoretic approaches to the transport
of non-Fermi liquids \cite{barkeshli,raghu,patel}. As in holography, these studies focused on the bottleneck associated with the relaxation
of momentum, instead of the traditional Boltzmann route of analyzing the remnants of quasiparticles.  
In many cases, it was found that the dominant momentum relaxation mechanism was associated
with the coupling of neutral, bosonic modes to random impurities, and that the influence of the charged fermionic excitations
near the Fermi surface was not as important. Furthermore, the bosonic contribution led to resistivities similar to those
obtained in the holographic theory. These features validate use of the holographic approach, which omits Fermi surface effects in the classical gravitational theory.

The purpose of the present paper is to extend the holographic result in Eq.~(\ref{rhomain}) to higher temperatures,
until there is eventually a crossover to an ultraviolet (UV) CFT with $z=1$. To this end, we shall obtain explicit geometries which
contain a crossover from a AdS$_4$ regime in the UV, to a hyperscaling violating metric in IR, ending at the horizon
of a black brane induced by the non-zero temperature. This is motivated by condensed matter studies 
of quantum phase transitions in metals with weak Landau damping \cite{raghu,sokol,georges,fitz1,fitz2,aass14};
in such situations, there can be a significant intermediate temperature regime where the quantum criticality
is described by that of a $z=1$ CFT, along with the usual Landau-damped non-Fermi liquid regime at the lowest
$T$. The holographic approach provides a solvable model for such a crossover between the non-Fermi liquid and $z=1$ regimes.

We begin in Section~\ref{sec:bh} by describing new methods for constructing the geometry of the crossover between a UV AdS$_{d+2}$
and an IR hyperscaling violating regime at a non-zero temperature; many technical details appear in Appendix~\ref{bhapp}.
This geometry describes the crossover induced by a relevant chemical potential $\mu$ on a CFT; however, we will express the crossover using the net charge density $\mathcal{Q}$ as a scale, rather than $\mu$.
Section~\ref{sec:rho} turns to a computation of the resistivity in the presence of weak disorder, following the analysis
of \cite{Lucas}. Finally, in Section~\ref{sec:sc} we address the instability of these geometries (without weak disorder) to superconductivity.

\section{Geometry of Crossover}
\label{sec:bh}

The tool that we will use in the computation of $\rho_{\mathrm{dc}}$ in this paper is gauge-gravity duality \cite{review1, review2, review3}:  this allows us to convert computations in a (large matrix $N$) strongly-coupled quantum field theory (QFT) into classical computations in gravity in one higher dimension.   The background metric in this higher dimensional spacetime encodes basic information about the physics in the  QFT.   For example, if we assume that the QFT is in flat space, and in a static and homogeneous state, the most general possible metric is \begin{equation}
\mathrm{d}s^2 = g_{MN}\mathrm{d}x^M\mathrm{d}x^N = \frac{L^2}{r^2}\left[\frac{a(r)}{b(r)} \mathrm{d}r^2 - a(r)b(r) \mathrm{d}t^2 + \mathrm{d}\mathbf{x}^2\right]  \label{intrometric}
\end{equation}The radius $r$ at which physics occurs in the bulk roughly maps on to the energy scale of the process in the boundary theory.   In particular, when this geometry has a (planar) black hole, the Hawking temperature $T$ -- related to the value of $r$ at the horizon -- the dual field theory is in a thermal state at the same temperature $T$.

There are special classes of geometries where $a(r)$ and $b(r)$ are simple power laws in $r$.  When these theories are properly coupled to bulk matter, they are dual to a QFT with scale-invariant spatial or temporal correlation functions \cite{Lucas}.  Such geometries are called hyperscaling violating geometries: if we express the power laws in $a$ and $b$ as \begin{subequations}\begin{align}
a(r) &\sim r^{-(d(z-1)-\theta)/(d-\theta)}, \\
b(r) &\sim r^{-(d(z-1)+\theta)/(d-\theta)},
\end{align}\end{subequations} then $z$ is equivalent to the dynamic critical exponent of the dual QFT, and $\theta$ is the hyperscaling violating exponent \cite{charmousis, iizuka, lsb, dong}.\footnote{In this paper, we will always take $\theta \le d-1$ and $z\ge 1+\theta/d$; the former is implied by the area law of entanglement (up to log-corrections), and the latter is then implied by the well-behavedness of the holographic dual.}   Of interest in this paper are geometries that interpolate between two of these regimes.   In particular, we will always consider the case where the UV theory ($r\rightarrow 0$ in these coordinates) is conformal ($z=1$, $\theta=0$).  The UV geometry is asymptotically anti-de Sitter (AdS) space.

Constructing such a geometry at a single temperature is straightforward \cite{edgar2011, gonzalez}.   It is harder to find an entire family of interpolating finite $T$ geometries without -- at every temperature $T$ -- re-solving the Einstein-matter system numerically.   The main difficulty that arises is to ensure that the couplings in the action are independent of $T$.  The only examples we have found in the literature occur at $z=\infty$ and for special values of $\theta$ \cite{rocha, gouteraux}; these are often derived by truncating solutions of supergravity via dimensional reduction \cite{cvetic}.     The main technical tool we will develop within holography is a bottom-up method to generate a one-parameter family (characterized by Hawking temperature) of static, homogeneous (in the boundary theory spatial directions) black holes ``semi-analytically" for arbitrary $z$ and $\theta$.  

The matter content we use to support these interpolating geometries is Einstein-Maxwell-dilaton (EMD) theory with two gauge fields.   The action of EMD theory with two Maxwell fields is \begin{equation}
S_{\mathrm{EMD}} = \int\mathrm{d}^{d+2}x\sqrt{-g}\left(\frac{1}{2\kappa^2}\left(R-2(\partial_M\Phi)^2 - \frac{V(\Phi)}{L^2}\right) - \frac{Z(\Phi)}{4e^2} F_{RS}F^{RS}- \frac{\tilde{Z}(\Phi)}{4e^2} \tilde{F}_{RS}\tilde{F}^{RS}\right).  \label{semd2}
\end{equation}
So as not to distract the non-specialist, we have left all details of how we construct families of finite temperature black holes from this action to Appendix \ref{bhapp}.   We state only the summary of our method here.   The static, homogeneous Einstein's equations are linear (in many gauges, such as Eq. (\ref{intrometric})) in a special metric degree of freedom, the ``emblackening factor" (EF) $b$, as are Maxwell's equations.   Since the EF does not arise in Maxwell's equations, we identify a scaling symmetry under which we can scale the EF and the $\tilde{F}$ Maxwell flux simultaneously, while maintaining a solution to the fully nonlinear equations of motion.   We then choose an EF which is a sum of two terms, one sourced by $F$ and one by $\tilde{F}$;  the one parameter family of black holes generated by scaling the $\tilde{F}$-sector of the theory corresponds to a family of black holes at different temperatures.    The computation of black holes at an infinite number of temperatures is reduced to solving a fixed number of 9 ordinary differential equations or constraint equations which have a highly nested structure.  Though we do not give exact analytic solutions to these equations,  they can be solved numerically using only integration, without any need to numerically solve differential equations using a typical finite difference or pseudospectral method.

This method gives us good analytic control over the resulting theory.  For example, in many cases we can directly show that thermodynamic properties of the dual theory, such as energy density $\epsilon$, pressure $P$, and chemical potential $\mu$ (associated to the charge $\mathcal{Q}$) take the expected form\footnote{Formally, we have only performed computations when there is a special relation between the charge density $\tilde{\mathcal{Q}}$ of the second charge and the temperature.  The thermodynamic potentials of the boundary theory also depend on $\tilde{\mathcal{Q}}$, and so these are the potentials restricted to a subclass of allowed equilibria.} \begin{subequations}\label{thermoeq1}\begin{align}
\epsilon(\mathcal{Q}, T) &= \left\lbrace \begin{array}{ll} \epsilon_1 \mathcal{Q}^{1+1/d} + A_d T^{d+1} &\ T\gg \mathcal{Q}^{1/d} \\  \epsilon_1 \mathcal{Q}^{1+1/d} +   \epsilon_2 T^{1+(d-\theta)/z} \mathcal{Q}^{1-(1-\theta/d)/z} &\ T \ll \mathcal{Q}^{1/d} \end{array} \right., \\
P(\mathcal{Q}, T) &= \left\lbrace \begin{array}{ll} P_1 \mathcal{Q}^{1+1/d} + d^{-1} A_d T^{d+1} &\ T\gg \mathcal{Q}^{1/d} \\ P_1 \mathcal{Q}^{1+1/d} + d^{-1} \epsilon_2 T^{1+(d-\theta)/z} \mathcal{Q}^{1-(1-\theta/d)/z}  &\ T \ll \mathcal{Q}^{1/d} \end{array} \right., \\
\mu(\mathcal{Q}, T) &= \left\lbrace \begin{array}{ll} \mu_0 &\ T\gg \mathcal{Q}^{1/d} \\  \mu_0 - \mu_1 T^{(1+(d-\theta)/z)/2}\mathcal{Q}^{(1-(d-\theta)/z)/2d} &\ T \ll \mathcal{Q}^{1/d} \end{array} \right..
\end{align}\end{subequations}with all of the coefficients in the above expressions dimensionless, positive, theory specific constants -- save for $A_d$, which is universal and identical to that of the AdS-Schwarzchild black hole. We describe how to extract them from the geometry in the appendix.   As expected, the charge density sets the temperature scale $T\sim \mathcal{Q}^{1/d}$ at which the geometry transitions from AdS to the hyperscaling-violating geometry -- this may not be desirable, but we explain why seems to always be the case in the appendix.

By gauge-gravity duality, the existence of a new gauge field in the bulk implies that there is another global conserved U(1) charge.   We do not provide any interpretation for this second charge, but note that this mystery charged sector will completely decouple from all computations of boundary theory quantities in this paper (though there are other computations where this sector will couple).   We also note that from a stringy perspective, dimensionally reducing supergravity from 10 or 11 spacetime dimensions naturally leads to a very large number of gauge fields in the low energy dimensionally reduced theory.

Of course, interpolating geometries can be found numerically within EMD theory alone, though the computation must be redone for every temperature.   We comment more on the extent to which our solutions differ from solutions to the pure EMD theory in the appendix.  In any case, our semi-analytic approach provides some insight into the structure of these black holes and is complementary to the standard numerical methods.

Moving beyond this, procedures for designing static and homogeneous black holes, similar to those described above, may also be fruitful in finding semi-analytically black holes with Bianchi horizons \cite{kachru2},\footnote{Some simple steps along this direction were taken in \cite{kachru1}, though the matter content supporting these geometries is unknown.}  which have shown up in holographic metal-insulator transitions \cite{hartnollnature, dgk}.

\section{Resistivity with Disorder}
\label{sec:rho}

We are now ready to use results from \cite{Lucas} to calculate the direct current conductivity, $\sigma_{\mathrm{dc}}$, of the dual theory to this family of EMD black holes, for any temperature $T$.    We begin below with some review of the topic, and then compare analytical predictions to numerical results for the family of theories holographically dual to interpolating geometries.   We conclude with extensions of the results of \cite{Lucas}. 

\subsection{Hydrodynamics}
 In general, our perturbative computation is in the regime described by momentum-relaxing hydrodynamics.   Let us briefly review the computation of $\sigma_{\mathrm{dc}}$ in this framework.   For a relativistic quantum field theory at finite temperature, such as we are considering, the equations of momentum-relaxing hydrodynamics are \cite{hkms} \begin{subequations}\begin{align}
\partial_\mu \langle T^{t\mu}\rangle &= F^{t\mu}\langle J_\mu\rangle, \\
\partial_\mu \langle T^{i\mu}\rangle &= -\frac{1}{\tau}\langle T^{i\mu}\rangle + F^{i\mu}\langle J_\mu\rangle , \\
\partial_\mu \langle J^\mu\rangle &= 0 .
\end{align}\end{subequations}with $F^{\mu\nu}$ the \emph{external} applied fields.   Applying a time and space independent, small electric field $\delta E_x$ in the $x$-direction, and using the generic form of the relativistic constituent equations of perfect hydrodynamics:\footnote{$F_{\mu\nu}$ should be thought of as a first derivative term, and so formally speaking we should include all first derivative terms in the gradient expansion below.   However as we are only interested in the conductivity at vanishing frequency and momentum, this will not affect the answer.} \begin{subequations}\begin{align}
\langle T^{\mu\nu}\rangle &=  \epsilon u^\mu u^\nu + P\left(\eta^{\mu\nu}+u^\mu u^\nu\right), \\
\langle J^\mu\rangle &=  \mathcal{Q} u^\mu + \sigma_{\mathrm{Q}} F^{\mu\nu}u_\nu,
\end{align}\end{subequations}with $\epsilon$ the energy density, $P$ the pressure, $\mathcal{Q}$ the charge density, and $\sigma_{\mathrm{Q}}$ a ``quantum critical" conductivity, we obtain \begin{equation}
\mathcal{Q} \delta E_x = \frac{1}{\tau} \langle T^{tx}\rangle \approx \frac{\epsilon + P}{\tau} \delta v_x = \frac{\epsilon + P}{\tau} \frac{\delta \langle J^x\rangle - \sigma_{\mathrm{Q}} \delta E_x}{\mathcal{Q}}.
\end{equation}By definition, $\sigma_{\mathrm{dc}}=\delta J_x/\delta E_x$, so we find \begin{equation}
\sigma_{\mathrm{dc}} = \sigma_{\mathrm{Q}} + \frac{\mathcal{Q}^2\tau}{\epsilon + P}.
\end{equation}

In holography, what we are essentially doing below by computing $\sigma_{\mathrm{dc}}$ is computing $\tau$ and $\sigma_{\mathrm{Q}}$.   A wealth of hydrodynamic relations thus allow us to compute thermoelectric transport \cite{hkms, thermoel1, thermoel2, thermoel3} as well as transport in a magnetic field \cite{hkms, blakedonos}.   So for simplicity, we will focus on the computation of $\sigma_{\mathrm{dc}}$ henceforth.

\subsection{Disorder in Holography}
As in \cite{Lucas}, we add disorder by coupling a time-independent source field, which is a Gaussian random function at each position in space, to an operator $\mathcal{O}$. Holographically, the operator $\mathcal{O}$ is dual to a bulk field $\psi$, and has the action
\begin{equation}
S_\psi = - \frac{1}{2} \int \mathrm{d}^{d+2}x\sqrt{-g} \left(\partial_M \psi \partial^M \psi + B(\Phi) \psi^2\right), \label{spsi}
\end{equation}  
along with a boundary source term associated with the random field.
The key result that we will use is the following formula, derived in \cite{Lucas}: \begin{equation}
\sigma_{\mathrm{dc}} = \frac{L^{d-2} Z(\Phi(r_{\mathrm{h}}))}{e^2r_{\mathrm{h}}^{d-2}} + \frac{d\mathcal{Q}^2r_{\mathrm{h}}^d}{L^d} \left(\int \frac{\mathrm{d}^d\mathbf{k}}{(2\pi)^d}k^2\psi(\mathbf{k},r_{\mathrm{h}})^2\right)^{-1}.  \label{eqsigmadc}
\end{equation}
Here $r_{\mathrm{h}}$ is the location of the black hole horizon;  it is the value of $r$ where the emblackening factor $b(r_{\mathrm{h}})=0$.  Similar results have been found for periodic potentials \cite{blake2} or translational symmetry breaking axions \cite{gouteraux2}.   There are encouraging signs that a (slightly more complicated) formula may even hold when the full nonlinear backreaction of $\psi$ is taken into account \cite{donos1409}.  The first term in Eq. (\ref{eqsigmadc}) corresponds to the conductivity due to pair creation of particles/antiparticles ($\sigma_{\mathrm{Q}}$ in hydrodynamics), and the latter corresponds to scattering off of disorder ($\mathcal{Q}^2\tau/(\epsilon+P)$ in hydrodynamics).   We provide a sketch of the derivation of this formula in Appendix \ref{rhodcappendix}.

It is actually slightly subtle that Eq. (\ref{eqsigmadc}) holds for our black holes, which have two gauge fields present.    This means that the boundary theory has two conserved currents, and therefore one must talk about a conductivity \emph{matrix}.   This matrix will contain three independent elements:  $\sigma_{JJ}$, $\sigma_{J\tilde{J}}$, and $\sigma_{\tilde{J}\tilde{J}}$.   Remarkably, it turns out that the computation of $\sigma_{JJ}(\omega=0)$ is insensitive to the presence of the second conserved current, in holography -- the formula Eq. (\ref{eqsigmadc}) holds when there are either one or two gauge fields.  Much of the previous work on transport in disordered strange metals actually computes the direct curent resistivity, which (for a system with a single conserved current) is given by \begin{equation}
\rho_{\mathrm{dc}} \equiv \frac{1}{\sigma_{\mathrm{dc}}}.   \label{eq12}
\end{equation}For our theories, this is not necessarily true -- $\rho_{JJ}$ is a matrix element which depends on the off-diagonal $\sigma_{J\tilde{J}}$ conductivity.    However, for the purposes of comparing with previous literature, we will simply define the quantity $\rho_{\mathrm{dc}}$ via Eq. (\ref{eq12}).   This quantity is more universal and will be our focus.   For simplicity, we will refer to it as a ``resistivity" below, but one should keep in mind that formally the holographic resistivity may be different.   

In particular, as we detail in Appendix \ref{bhapp}, our exact black hole constructions rely on a peculiar scaling between the charge density $\tilde{\mathcal{Q}}$ associated with the new bulk gauge field, and the temperature $T$.   While the form of Eq. (\ref{eqsigmadc}) for $\sigma_{JJ}$ is completely general and continues to hold for arbitrary choices of $\tilde{\mathcal{Q}}$, it is possible to obtain anomalous scaling in $\rho_{JJ}$.   However, in both the low $T$ and high $T$ regimes -- where we will show the scaling in Eq. (\ref{rhomain}) holds -- our metrics are parametrically good approximate solutions of Einstein's equations with $\tilde{\mathcal{Q}}=0$, where Eq. (\ref{eq12}) becomes an identity.    Even in the ``intermediate" $T$ regime, it is often the case that our metric and dilaton fields (and thus the neutral scalar background) approximately satisfiy their equations of motion with the tilde gauge fields identically set to vanish.    Finally, we emphasize that Eq. (\ref{eqsigmadc}) is independent of the precise way that we tune $\tilde{\mathcal{Q}}$ to generate families of solutions -- it holds for all static, homogeneous, asymptotically AdS geometries dual to theories with two conserved charges.    The conductivities we are computing are therefore unlikely to have any spurious behavior sensitive to our particular geometric construction.

In order to compute $\rho_{\mathrm{dc}}$, we need a reasonable ansatz for $B(\Phi)$.   We make the following choice: \begin{equation}
B(\Phi) \approx \left\lbrace \begin{array}{ll} \Delta_{\mathrm{UV}}(\Delta_{\mathrm{UV}}-d-1)L^{-2} &\ \Phi \ll 1 \\   B_0 G_0^{-1} L^{-2}r^{-2\theta/(d-\theta)}    &\   \Phi \gg 1 \end{array}\right..   \label{Bphieq}
\end{equation}where $\Delta_{\mathrm{UV}}$ is the UV dimension of the dual operator to $\psi$, and the constant $G_0 \approx g_{rr} r^{-2\theta/(d-\theta)}$ when $r\gg \mathcal{Q}^{1/d}$.   $B_0$ will control the effective dimension of the operator in the IR, $\Delta_{\mathrm{IR}}$, according to the formula \begin{equation}
B_0 = \left(\Delta_{\mathrm{IR}} - \frac{d+z}{2}\right)^2 - \left(\frac{d-\theta+z}{2}\right)^2.
\end{equation}  
As the critical exponents of the theory are changing, we expect that in general $\Delta_{\mathrm{UV}} \ne \Delta_{\mathrm{IR}}$.   Finally, we add disorder by imposing a weak background of scalar hair on top of our EMD black hole.  We can compute $\rho_{\mathrm{dc}}$ without worrying about the back-reaction on the geometry so long as the disorder strength is small.   To add disorder, we simply choose the boundary conditions so that as $r\rightarrow 0$, \begin{equation}
\psi(\mathbf{k},r) \sim g(\mathbf{k})r^{d+1-\Delta_{\mathrm{UV}}}  \label{psias}
\end{equation}where we assume that $g(\mathbf{k})$ are independent, identically distributed Gaussian random variables: \begin{subequations}\begin{align}
\mathbb{E}[g(\mathbf{k})] &= 0,\\
\mathbb{E}[g(\mathbf{k})g(\mathbf{q})] &= \varepsilon^2 \delta(\mathbf{k}+\mathbf{q}).
\end{align}\end{subequations}We have used $\mathbb{E}[\cdots]$ to denote averages over the quenched random-field disorder.  

It is crucial that the form of Eq. (\ref{Bphieq}) holds in order to reproduce Eq. (\ref{rhomain}) in the IR, a result which can be verified using an independent memory matrix approach \cite{Lucas}.  If this is not obeyed, then the AdS radius $L$ becomes a physical length scale, measurable in correlation functions, as was the case in \cite{dong}.   There is, as of yet, no good understanding of why this should be true generally in holography;  perhaps studying supergravity truncations further can shed light on this issue.

We assume that $\Delta_{\mathrm{UV}} < (d+2)/2$, which is the Harris criterion \cite{herzog} for a CFT (implying that disorder can be treated perturbatively).   We also choose that $d+1-\Delta_{\mathrm{UV}} < \Delta_{\mathrm{UV}}$ and is the dominant power in the UV.   Thus indeed the UV asymptotics in Eq. (\ref{psias}) is correct.    It is important that the Harris criterion be satisfied;  if it is not, then there are non-perturbative UV corrections to the theory.   This can be seen holographically by noting that the back-reaction from $\psi$ on the geometry will blow up as $r\rightarrow 0$.   A similar Harris criterion exists for a hyperscaling violating theory \cite{Lucas}, and we will take $\Delta_{\mathrm{IR}}$ to satisfy this criterion as well:\begin{equation}
\frac{d+z}{2}< \Delta_{\mathrm{IR}} < \frac{d-\theta}{2}+z.   \label{harriseq}
\end{equation}

Our computation of $\rho_{\mathrm{dc}}$ is reliable at every temperature $T$ for which $\psi$ does not strongly backreact on the geometry.   A holographic analysis shows that this approximation is valid so long as $T\gg T_{\mathrm{np}}$, with \begin{equation}
T_{\mathrm{np}} \sim \hat{\mathcal{Q}}^{1/d}  \left(\frac{\kappa^2\varepsilon^2}{\hat{\mathcal{Q}}^{(d+2-\Delta_{\mathrm{UV}})/d}}\right)^{2z/(2z-2\Delta_{\mathrm{IR}}+d-\theta)}
\end{equation}where\footnote{Note that $\mathcal{Q}$ and $\hat{\mathcal{Q}}$ have the same dimensions.}\begin{equation}
\hat{\mathcal{Q}} \equiv \mathcal{Q} \frac{e\kappa}{L^{d -1}}.   \label{qhat}
\end{equation}For $T\lesssim T_{\mathrm{np}}$, the backreaction of the scalar cannot be ignored, non-perturbative effects become important, and our calculation needs to be modified to account for this.

We have assumed in this expression that $\varepsilon\rightarrow 0$, so that disorder may be treated perturbatively at temperatures low enough that the geometry is approximately hyperscaling violating near the horizon.   This means that we require $T_{\mathrm{np}} \ll \hat{\mathcal{Q}}^{1/d}$.    Note that as $\varepsilon\rightarrow 0$, this implies the Harris criterion in the IR must be satisfied.

To compute $\rho_{\mathrm{dc}} \equiv 1/\sigma_{\mathrm{dc}}$, we thus need to determine the value of $\int \mathrm{d}^d\mathbf{k}\; k^2\psi(r_{\mathrm{h}})^2$.   This can be estimated analytically by a matched asymptotic analysis -- see Appendix \ref{rhodcappendix} -- and from Eq. (\ref{eqsigmadc}) we find the following result: \begin{equation}
\rho_{\mathrm{dc}} \sim \left\lbrace \begin{array}{ll} e^2 L^{2-d} T^{d-2} &\ T \gg T_{\mathrm{pc}} \\ \varepsilon^2 \mathcal{Q}^{-2} T^{2\Delta_{\mathrm{UV}}} &\ \hat{\mathcal{Q}}^{1/d}\ll T \ll T_{\mathrm{pc}} \\  \varepsilon^2 \mathcal{Q}^{-2}\hat{\mathcal{Q}}^{2\Delta_{\mathrm{UV}}/d} (T\hat{\mathcal{Q}}^{-1/d})^{2(1+\Delta_{\mathrm{IR}}-z)/z} &\  T_{\mathrm{np}} \ll T \ll \hat{\mathcal{Q}}^{1/d}   \end{array}\right.  \label{rhosketch}
\end{equation}where \begin{equation}
T_{\mathrm{pc}} \sim \left( \frac{e L^{1-d/2}\mathcal{Q}}{\varepsilon}\right)^{2/(2+2\Delta_{\mathrm{UV}}-d)}.  \label{tpc}
\end{equation} 
The results in the 2 regimes with $T \ll T_{\mathrm{pc}}$ agree with Eq.~(\ref{rhomain}), after accounting for the $z=1$ exponent
in the UV CFT regime.    The result for $T\gtrsim T_{\mathrm{pc}}$ follows from the fact that at very large temperatures, the first term in Eq. (\ref{eqsigmadc}) dominates the electrical conductivity.   

This result can be understood on simple physical grounds.   As $T\rightarrow\infty$, the conductivity is dominated by the spontaneous excitation of particle/antiparticle pairs in the CFT.   As these particles have opposite charge, they contribute to a net electric current by moving in opposing directions (while contributing no net momentum).   As these excitations do not carry a net momentum, they are unaffected by disorder, which becomes increasingly efficient at dissipating the momentum carried by the motion of the background charge density $\mathcal{Q}$.    In fact, in this regime the scaling in Eq. (\ref{rhosketch}) matches that of a charge-neutral conformal plasma.    At weak disorder but finite charge density, momentum relaxation is a very slow process and thus, below a \emph{parametrically large} temperature $T_{\mathrm{pc}} \sim \varepsilon^{-\#}$ (the ``pair creation" temperature), the non-zero overlap between momentum and electric current controls the rate at which the electric current can decay.  For $T\ll T_{\mathrm{pc}}$,  it is the slow decay of the momentum associated with the background charge density $\mathcal{Q}$ which controls the resistivity.  The temperature scale $\hat{\mathcal{Q}}^{1/d}$ controls the transition from the CFT to the low energy effective hyperscaling violating theory.   Whether the theory is approximately a CFT or a theory with finite $z$ and $\theta$ at temperature $T$ then also qualitatively changes the $T$-scaling of $\rho_{\mathrm{dc}}$.    Finally, as mentioned above, our perturbative result breaks down at $T_{\mathrm{np}}$, when the disorder is no longer a weak perturbation to the strange metal.   There are no reliable techniques as of yet for extending the computation of $\rho_{\mathrm{dc}}$ to lower temperatures -- the qualitative behavior of $\rho_{\mathrm{dc}}$ in this regime is an open question.

\begin{figure}
\centering
\includegraphics[width=4.5in]{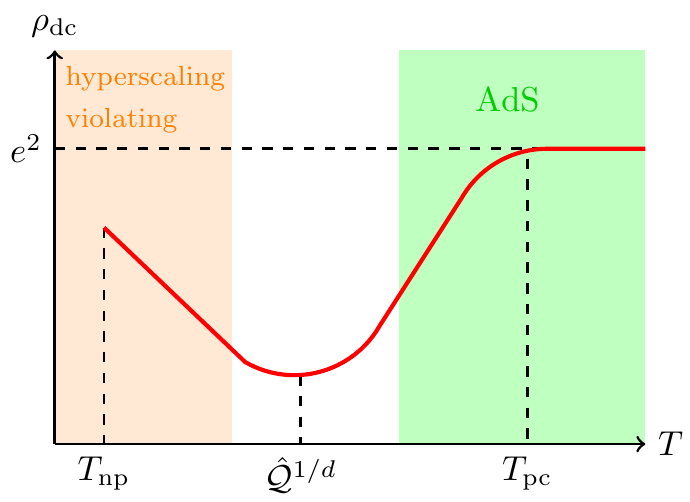}
\caption{A sketch of $\rho_{\mathrm{dc}}$ in $d=2$, showing $\rho_{\mathrm{dc}}(T)$, as well as the portion of the geometry capturing the processes dominating the resistivity.}
\label{cartoonfig}
\end{figure}

Note that it is possible that $\rho_{\mathrm{dc}}$ reaches a minimum at a finite temperature $T\sim \hat{\mathcal{Q}}^{1/d}$.  A cartoon of Eq. (\ref{rhosketch}) appears in Figure \ref{cartoonfig}.   We shortly verify numerically that this can indeed happen in Figure \ref{3dplot}.

We should keep in mind, of course, that the computation may not be reliable up to $T_{\mathrm{pc}}$ simply because $T_{\mathrm{pc}} \sim \varepsilon^{-\#}$, and the approximate description of the metal as a CFT in a continuous space may not be valid at arbitrarily high $T$. 

\subsection{Numerical Results}
Once we explicitly construct an interpolating geometry using the techniques described earlier, a computation of $\rho_{\mathrm{dc}}$ reduces to a computation of $\psi(\mathbf{k},r_{\mathrm{h}})$, given the random boundary condition Eq. (\ref{psias}), along with the boundary condition that $\psi$ is finite at the horizon.

The computation of $\psi(\mathbf{k},r_{\mathrm{h}})$ is, luckily quite simple.  Given our numerically integrated black hole geometries, we need only solve the wave equation \begin{equation}
\frac{r^{d+2}}{a}\left(\frac{b}{r^d}\psi^\prime\right)^\prime + k^2 r^2 \psi  = L^2 B(\Phi(r))\psi   \label{psiwave}
\end{equation}
for a solution obeying our two boundary conditions.  In practice, it is easiest to enforce that the solution be finite at the black hole horizon by choosing initial conditions such that $\tilde \psi(r_{\mathrm{h}}-\delta)=1$ and $\tilde \psi^\prime(r_{\mathrm{h}}-\delta)=\text{finite}$ (here $\delta>0$ is an infinitesimal numerical parameter).\footnote{One can do an asymptotic analysis and determine the exact value of $\tilde \psi^\prime(r_{\mathrm{h}})$, given $\tilde \psi(r_{\mathrm{h}})$, but in practice since there is a divergent mode at the horizon, picking any finite constant, and taking the limit where $\delta\rightarrow 0$, recovers the same answer.  We have done the calculation by both doing the series expansion and by simply setting $\tilde \psi^\prime(r_{\mathrm{h}}) = 0$ and the answers are within the numerical error obtained by integrating over momentum modes.}    It is straightforward to see that \begin{equation}
\psi(\mathbf{k}, r_{\mathrm{h}}) \approx g(\mathbf{k}) \frac{\delta^{d+1-\Delta_{\mathrm{UV}}}}{\tilde\psi(k, \delta)},
\end{equation}
which, along with breaking up $\int \mathrm{d}|\mathbf{k}|$ with a simple Riemann sum, gives a very straightforward algorithm to compute $\int \mathrm{d}^d\mathbf{k} \; k^2\psi^2$.

\begin{figure}
\centering
\includegraphics[width=6in]{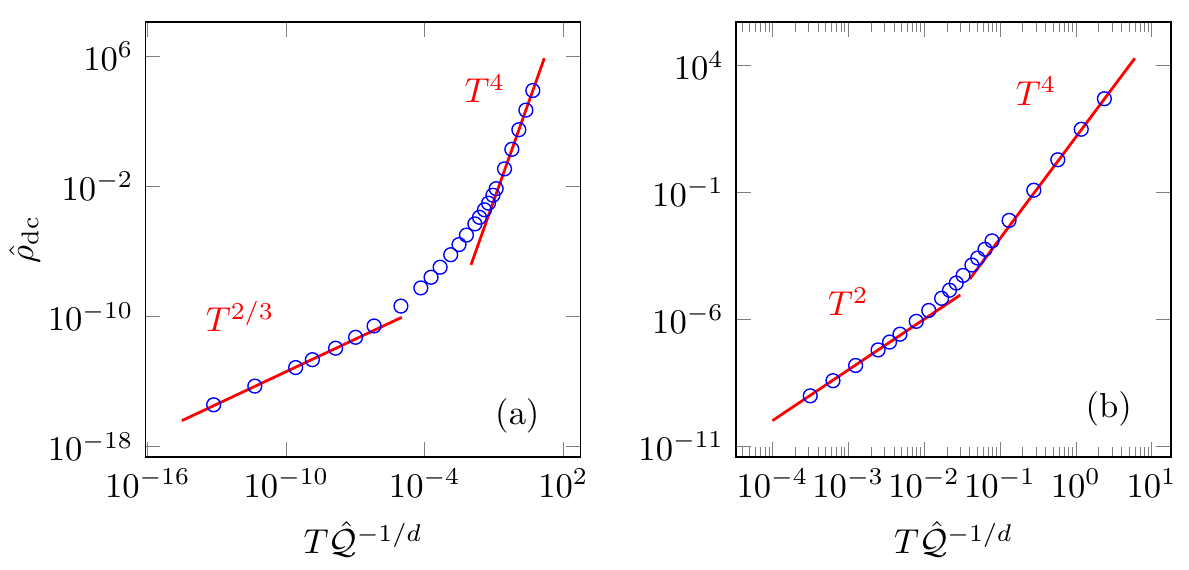}
\caption{We show $\hat\rho_{\mathrm{dc}}$ vs. $T$ for the $d=2$ geometries with (a) $z=3$, $\theta=1$, (b) $z=2$, $\theta=-2$.  In both cases $\Delta_{\mathrm{UV}}=2$  and $\Delta_{\mathrm{IR}}=3$.  The solid lines are fits to analytical predictions (up to overall coefficients), which are computed using Eq. (\ref{rhosketch}).   Error in numerical methods is substantially smaller than the data markers.  Note that $\Delta_{\mathrm{UV}}$ is ``marginal" from the Harris criterion; though at very large $T \gg T_{\mathrm{pc}}$ the geometry may deviate from AdS logarithmically due to the presence of disorder, such an effect is well beyond our regime of validity.   This choice of $\Delta_{\mathrm{UV}}$ appeared in \cite{patel} as well.}
\label{2dplot}
\end{figure}

Our numerical results are shown in Figure \ref{2dplot}, when $z$ is finite. In all plots we show a dimensionless $\rho_{\mathrm{dc}}$ defined by \begin{equation}
\hat\rho_{\mathrm{dc}} = \frac{\mathcal{Q}^2\hat{\mathcal{Q}}^{-2\Delta_{\mathrm{UV}}/d}}{\varepsilon^2}\rho_{\mathrm{dc}}.
\end{equation}Formally we should write $\mathbb{E}[\rho_{\mathrm{dc}}]$ above instead, but we will argue at the end of this section that almost every sample will (at leading order) have the same resistivity when disorder is weak.  We only consider temperatures below $T_{\mathrm{pc}}$.   It is evident that our scaling theory is quite accurate, though in some cases there can be a rather large range of temperatures in the transition regime between the two limiting scaling regimes.

In the case where $z\rightarrow\infty$, and $\theta/z = -\eta$ is held fixed \cite{shaghoulian2}, the story is a bit more subtle holographically.  As we explain in Appendix \ref{neuapp}, the effective IR scaling dimension of operators becomes momentum-dependent.   This slightly complicates the previous asymptotic analysis, and we find logarithmic corrections to $\rho_{\mathrm{dc}}$: \begin{equation}
\rho_{\mathrm{dc}} \sim \frac{\varepsilon^2 T^{2(\Delta_{\mathrm{IR}}/z-1)}}{(\log (1/r_{\mathrm{HV}}T))^{1+d/2}}, \;\;\; T_{\mathrm{np}} \ll T \ll \hat{\mathcal{Q}}^{1/d}.  \label{rholog}
\end{equation}

\begin{figure}
\centering
\includegraphics[width=6.6in]{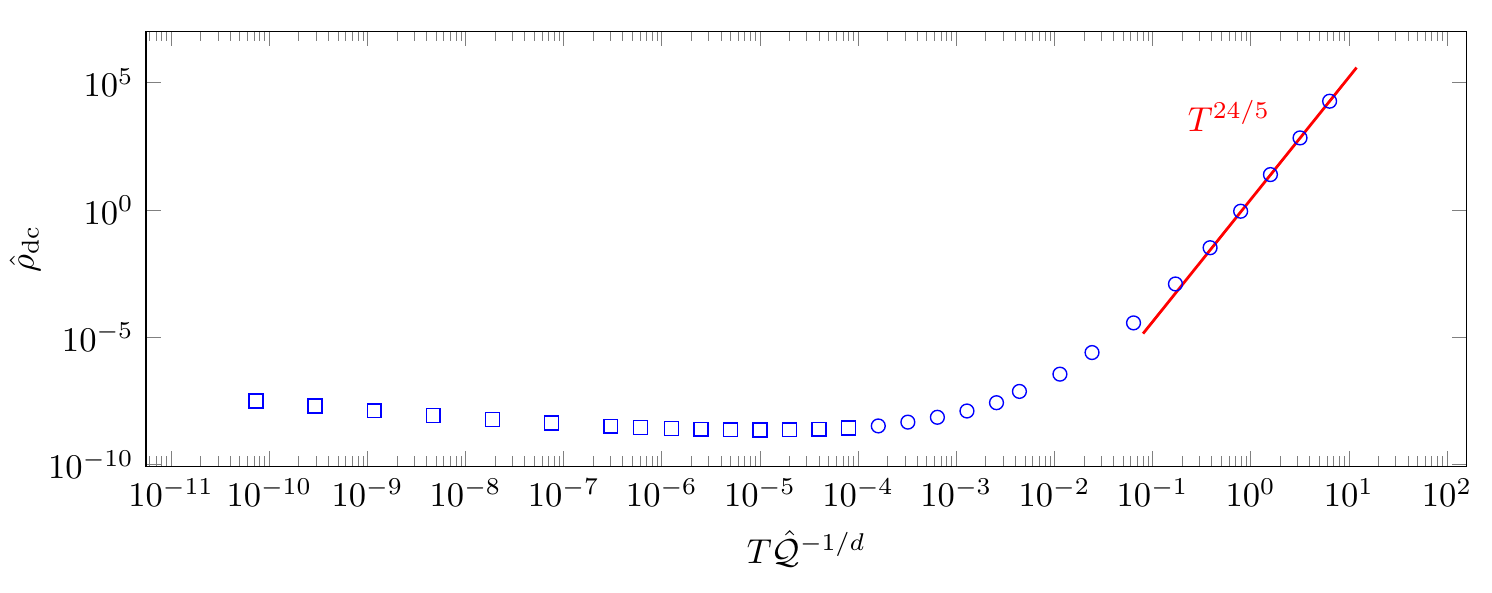}
\caption{We show $\hat\rho_{\mathrm{dc}}$ vs. $T$ for the $d=3$ geometry with $\eta=3$.  We took $\Delta_{\mathrm{UV}} = 12/5$ and $\Delta_{\mathrm{IR}} = 3z/4$.  Circular data points denote data where we have done a full numerical computation of the strength of the disordered scalar hair.   For square data points, we have used matched asymptotic expansions (see e.g. \cite{sera}) to extend the numerical methods over an additional 8 orders of magnitude in $T$.  Note that $\rho_{\mathrm{dc}}$ has a minimum for $T\hat{\mathcal{Q}}^{-1/d}\sim 10^{-5}$.   Error in numerical methods is substantially smaller than the data markers.}
\label{3dplot}
\end{figure}

Numerical results for the geometry with $d=3$, $z=\infty$, $\eta=3$ are shown in Figure \ref{3dplot}.  The transition into the log-corrected scaling regime given by Eq. (\ref{rholog}) appears highly delayed.   Because of the holographic nature of the disordered scalar profile (higher momentum modes decaying algebraically, as opposed to exponentially when $z<\infty$) this is not surprising.   Nonetheless, we are able to detect a minimum in $\rho_{\mathrm{dc}}(T)$ -- there exists a temperature below which disorder \emph{enhances} the scattering rate of momentum.

\subsection{Fluctuations}
Let us also briefly comment on the fluctuations of $\rho_{\mathrm{dc}}$.   The following discussion is \emph{completely independent} of holography and relies only on the general structure of the resistivity.     In holography, or via memory matrices \cite{hkms, dsz, raghu, Lucas, patel} we have found that the resistivity takes the form \begin{equation}
\rho_{\mathrm{dc}} = \int \mathrm{d}^d\mathbf{k}\mathrm{d}^d\mathbf{q} \; g(\mathbf{k})g(\mathbf{q})\mathcal{G}(\mathbf{k},\mathbf{q}).
\end{equation}In the memory matrix formalism, $\mathcal{G}$ is related to the small $\omega$ leading-order imaginary component of a Green's function; in holography, $\mathcal{G}$ is related to the scalar hair profile at the horizon of the black hole. 

Assuming Gaussian disorder for $g(\mathbf{k})$, it is straightforward to compute all sample-to-sample fluctuations in $\rho_{\mathrm{dc}}$.    Assuming a cubic sample of width $\mathcal{V}^{1/d}$ in each spatial dimension (where $\mathcal{V}$ is the total volume of the sample), with periodic boundary conditions for simplicity, we regularize the momentum integrals with sums: \begin{equation*}
\int \mathrm{d}^d\mathbf{k} \rightarrow \frac{(2\pi)^d}{\mathcal{V}}\sum_{\mathbf{k}}.
\end{equation*}We assume that $\mathcal{V}^{1/d}T^{1/z} \gg 1$, so that scattering events off of long-wavelength disorder  are not exponentially suppressed.     The number of modes which are relevant for the computation of resistivity scales as $\mathcal{V}T^{d/z}$.

For simplicity, we will only show explicitly the computation of $\mathrm{Var}(\rho_{\mathrm{dc}})$: \begin{equation}
\mathrm{Var}(\rho_{\mathrm{dc}}) = \mathbb{E}\left[\rho_{\mathrm{dc}}^2\right]  - \mathbb{E}[\rho_{\mathrm{dc}}]^2 
\end{equation}Using Wick's Theorem, this variance simplifies to \begin{equation}
\mathrm{Var}(\rho_{\mathrm{dc}}) = 2 \frac{(2\pi)^{2d}}{\mathcal{V}^2}\sum_{\mathbf{k},\mathbf{q}} \mathcal{G}_{\mathbf{k},\mathbf{q}}\mathcal{G}_{-\mathbf{k},-\mathbf{q}} \varepsilon^4
\end{equation}Now, $\mathcal{G}_{\mathbf{k},\mathbf{q}}$ is related to the imaginary part of a retarded Green's function in the \emph{translationally invariant} field theory (without disorder).   Therefore, only the diagonal modes where $\mathbf{k}+\mathbf{q}=\mathbf{0}$ are non-vanishing.   Further using that for $|\mathbf{k}|<T^{1/z}$, $\mathbf{G}\sim \mathcal{G}_T$ is approximately $\mathbf{k}$-independent, it is straightforward to conclude that \begin{equation}
\frac{\mathrm{Var}(\rho_{\mathrm{dc}})}{\mathbb{E}[\rho_{\mathrm{dc}}]^2} \sim \frac{1}{\mathcal{V}T^{d/z}}.  \label{rms}
\end{equation}

As we are in a ``high temperature" regime, the answer is what one would naively expect -- sample-to-sample fluctuations in $\rho_{\mathrm{dc}}$ are $1/\mathcal{N}$ suppressed, where $\mathcal{N} \sim \mathcal{V}T^{d/z}$ is the number of relevant disorder modes off of which momentum can scatter.   This can be contrasted with the non-self-averaging results found for free theories in the mesoscopic regime \cite{leestone, altshuler, altshuler2, imry, leestone2}.   For example, at low temperatures in $d=2$,  the variance of $\sigma_{\mathrm{dc}}$ should be a universal constant.    At higher temperatures, these systems transition to a regime analogous to Eq. (\ref{rms}) \cite{altshuler2, leestone2}.  

\section{Superconductivity}
\label{sec:sc}

Let us briefly turn to another question:  do we expect these holographic strange metals to be prone to superconductivity?  Recent non-holographic work that suggested that the Ising-nematic quantum critical point was always unstable to superconductivity at low $T$ \cite{max}.   In particular, the onset of superconductivity arose \emph{before} the IR scaling regime in $\rho_{\mathrm{dc}}$.    It is therefore of interest to see whether this instability is present in a large class of strongly-coupled theories, and in particular, whether this instability arises for temperatures all the way up to $T\sim \hat{\mathcal{Q}}^{1/d}$.   A priori there is no reason this needs to occur, because the calculation of \cite{max} relied explicitly on the presence of the Fermi surface.   

Note that we are assuming there is no quenched disorder for this section.   This superconductivity is purely a property of the interpolating geometries alone.

There is a very elegant holographic mechanism for both superconductivity and superfluidity, where one can heuristically understand that these effects arise because an effective mass term for a charged bulk scalar becomes too negative \cite{gubser, hhh1, hhh2, herzog2}, causing an instability whether the charged scalar condenses.   For other work on holographic superconductors with hyperscaling violation (but with a different action for the scalar), see \cite{salvio, fan}.  

We suppose that we now make our $\psi$ scalar charged under the Maxwell field, with charge $Q$ -- other than replacing $\partial_M$ with $\partial_M - \mathrm{i}QA_M$, we leave Eq. (\ref{spsi}) unchanged.   It is known from the theory of holographic superconductors that in the linear response regime, a perturbation $\psi$ will be real.   Assuming that we have a spatially homogeneous perturbation with time-dependence $\sim \mathrm{e}^{-\mathrm{i}\omega t}$,  in our coordinates:\footnote{In the notation of Appendix \ref{bhapp}, $A_t = p(r)-p(r_{\mathrm{h}})$, with $p$ the integral of $p^\prime$ with $p(\infty)\equiv 0$.} \begin{equation}
br^d\left(\frac{b}{r^d}\psi^\prime\right)^\prime = \left(\frac{ab}{r^2}B(\Phi) - Q^2 A_t^2\right)\psi - \omega^2\psi.
\end{equation}

A particularly simple task is to simply set $\omega=0$, and look for the critical value of $Q$ at which this equation first admits an equation with the appropriate boundary conditions \cite{gubser} -- namely, that the leading order asymptotic mode of $\psi$ in the UV vanishes, and that $\psi$ is finite at the horizon.    This does not \emph{prove} that there is an instability to a superconducting phase, but it is quite suggestive.   

\begin{figure}
\centering
\includegraphics[width=4in]{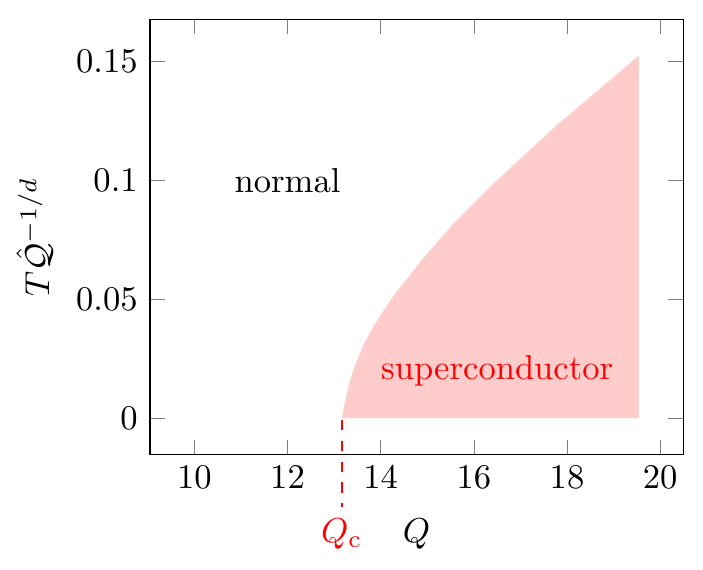}
\caption{We show the values of $Q$ and $T$ for which there is a likely instability to superconductivity for our previous $d=z=-\theta=2$ geometry.  The boundary between red and white defines where a marginal mode of the charged scalar exists.}
\label{scplot}
\end{figure}

We have performed this task numerically for the $d=2$, $z=-\theta=2$ geometry we presented earlier, by giving the scalar field $\psi$ (with the same $B(\Phi)$ used to compute $\rho_{\mathrm{dc}}$) a charge $Q$ as described above.   The results are shown in Figure \ref{scplot}.   The curious feature of this plot is that it appears as though there exists a $Q_{\mathrm{c}}$ ($\approx 13.16$ in this case), below which there is no instability at \emph{any} $T$.    When $Q\gtrsim Q_{\mathrm{c}}$, the critical temperature rapidly becomes comparable to the temperature at which the geometry transitions from AdS to the hyperscaling violating region.   Evidently, the superconductor arising from condensation of this charged scalar is associated with an instability of the transition region of the geometry.   Thus, for certain scalars which have a large enough charge, we indeed recover the basic phenomenology of \cite{max}.   In our case, it is also possible for no superconducting instability to exist, even with a charged scalar at $T=0$.   It would be interesting whether there is some universal mechanism underlying this curious phenomenology both in holographic and non-holographic models.    There may be other holographic models where superconductivity persists to $T=0$, depending on the nature of the dilaton coupling to $\psi$.

A simple explanation for the existence of this critical charge $Q_{\mathrm{c}}$ is as follows.   Recall that superconductivity can arise from an effective mass for the scalar $\psi$ which is ``too negative".   Deep in the IR, however, the uncharged ($B(\Phi)$) contribution to the effective mass is parametrically larger than the charged contribution ($-Q^2A_t^2$): \begin{equation}
\frac{ab}{r^2}B(\Phi) \sim r^{-2dz/(d-\theta)}, \;\;\;\; A_t^2 \sim r^{-2d-2dz/(d-\theta)}.
\end{equation}Thus, we expect if there is a region of the geometry where the mass becomes too negative, it must be \emph{before} the hyperscaling violating region of the geometry $r\ll \hat{\mathcal{Q}}^{-1/d}$.  This directly implies that such an instability would persist at least until temperatures $T\sim \hat{\mathcal{Q}}^{1/d}$, which is exactly what we numerically observe.

For large $Q$, we find $T_{\mathrm{c}}(Q) \sim Q^{1/2}$, consistent with mean-field superconductivity in the $\mathrm{AdS}_4$-Schwarzchild geometry \cite{hhh2}.   This regime occurs at very large $Q$ and is not visible in the figure.    For small $Q-Q_{\mathrm{c}}$, we estimate from numerics that $T_{\mathrm{c}}(Q) \sim Q-Q_{\mathrm{c}}$; there is no superconducting instability when $Q<Q_{\mathrm{c}}$.   Similar behavior appears for other choices of $B(\Phi)$.

A similar effect, where superconductivity requires a charge $Q$ larger than some critical $Q_{\mathrm{c}}$, has also recently been reported in some Q-lattice geometries \cite{yiling}.   

\section{Conclusions}

A significant feature of our results for the crossover from the UV CFT to the IR hyperscaling-violating strange metal is that 
the crossover always occurred at the length scale $\mathcal{Q}^{1/d}$, and associated energy scales. We were not able to find
parameter regimes in which the crossover occurred at scales parametrically different from $\mathcal{Q}^{1/d}$, as
explained in Appendix~\ref{bhapp}. This should be contrasted from the behavior in the non-Fermi liquid field 
theory \cite{sokol,georges,fitz1,fitz2,raghu,aass14}, where arguments have been made for a crossover from 
Landau-damped to $z=1$ physics at an energy scale significantly smaller
than the Fermi energy. It could be that the present holographic formulation is missing some essential ingredient, such as
the proximity to a Mott insulator \cite{georges}, and this is an important subject for future investigation.

A related notable feature of our results concerns the onset of superconductivity, as discussed in Section~\ref{sec:sc},
due to the condensation of a charged scalar. For a sufficiently massive scalar, or a scalar with a small enough charge, there was no condensation, and the IR 
hyperscaling-violating metal was stable at low $T$. On the other hand, in the regime in which the scalar did condense, its critical temperture rapidly became of order the energy scale associated with $\mathcal{Q}^{1/d}$. Thus almost all of the 
low $T$ non-Fermi liquid
regime was masked by superconductivity, as was also found in a recent field-theoretic analysis \cite{max}.

 It would be interesting to extend the holographic computation of resistivity into the strong-disorder regime.  A first step towards this goal has been taken numerically by constructing black holes with disordered scalar hair \cite{hartnollsantos}.   Whether or not strong disorder can overwhelm strong interactions to produce a highly inhomogeneous, ``localized" phase without transport ($\rho_{\mathrm{dc}}=\infty$) is an important goal for future works.

\addcontentsline{toc}{section}{Acknowledgements}
\section*{Acknowledgements}
We thank Richard Davison, Blaise Gout\'eraux and Koenraad Schalm for helpful discussions.   We also thank Pierre Le Doussal for suggesting the calculation of sample-to-sample fluctuations of $\rho_{\mathrm{dc}}$.
This research was supported by the NSF under Grant DMR-1360789, the Templeton foundation, and MURI grant W911NF-14-1-0003 from ARO.
Research at Perimeter Institute is supported by the Government of Canada through Industry Canada 
and by the Province of Ontario through the Ministry of Research and Innovation. 

\begin{appendix}
\titleformat{\section}
  {\gdef\sectionlabel{}
   \Large\bfseries\scshape}
  {\gdef\sectionlabel{\thesection. }}{0pt}
  {\begin{tikzpicture}[remember picture,overlay]
	\draw (-0.2, 0) node[right] {\textsf{Appendix \sectionlabel#1}};
	\draw[thick] (0, -0.4) -- (\textwidth, -0.4);
       \end{tikzpicture}
  }
\titlespacing*{\section}{0pt}{15pt}{20pt}

\section{Constructing Interpolating Black Holes}  \label{bhapp}
The equations of motion from the action Eq. (\ref{semd2}) are Einstein's equation: \begin{align}
R_{MN} - \frac{R}{2}g_{MN} &= \frac{\kappa^2}{e^2}\left(F_{MR}{F_N}^R - \frac{1}{4}F_{RS}F^{RS}g_{MN}\right) Z(\Phi) + \frac{\kappa^2}{e^2}\left(\tilde{F}_{MR}{\tilde{F}_N}^R - \frac{1}{4}\tilde{F}_{RS}\tilde{F}^{RS}g_{MN}\right) \tilde{Z}(\Phi) \notag \\
  &- \left(2(\partial_M\Phi)^2+\frac{V(\Phi)}{L^2}\right)\frac{g_{MN}}{2} + 2\partial_M \Phi \partial_N \Phi,
\end{align}Maxwell's equations \begin{equation}
\nabla_M \left(\tilde{Z}(\Phi)\tilde{F}^{MN}\right) = \nabla_M \left(Z(\Phi)F^{MN}\right) = 0,
\end{equation}and the dilaton equation of motion: \begin{equation}
4\nabla_M\nabla^M\Phi  = \frac{1}{L^2}\frac{\partial V}{\partial \Phi} + \frac{\kappa^2}{2e^2} F_{RS}F^{RS} \frac{\partial Z}{\partial \Phi}+ \frac{\kappa^2}{2e^2} \tilde{F}_{RS}\tilde{F}^{RS} \frac{\partial \tilde{Z}}{\partial \Phi}.
\end{equation} 

Using the metric ansatz Eq. (\ref{intrometric}), the gauge field ansatz \begin{equation}
F = p^\prime(r)\mathrm{d}r\wedge\mathrm{d}t, \;\; \tilde{F} = \tilde{p}^\prime(r)\mathrm{d}r\wedge\mathrm{d}t,
\end{equation} and the dilaton ansatz $\Phi = \Phi(r)$, we find that we can re-write Einstein's equations in the form: \begin{subequations}\begin{align}
2\Phi^{\prime 2} - \frac{\kappa^2}{e^2}  \frac{Z p^{\prime 2} + \tilde{Z}\tilde{p}^{\prime 2}}{ab}  \frac{r^2}{L^2}&= R_{rr} - \Xi g_{rr} =  R_{rr}  - \frac{\Xi L^2a}{r^2b}, \\
\frac{\kappa^2}{e^2} \frac{(Zp^{\prime 2} + \tilde{Z}\tilde{p}^{\prime 2}) b}{a} \frac{r^2}{L^2} &= R_{tt} - \Xi g_{tt} = R_{tt} + \frac{\Xi L^2 ab}{r^2}, \\
0 &= R_{ii} - \frac{\Xi L^2}{r^2} = \frac{rb^\prime - (d+1)b}{r^2a} - \frac{\Xi L^2}{r^2}.
\end{align}\end{subequations}The constant $\Xi$ -- the proportionality coefficient of $g_{MN}$ in Einstein's equations, is therefore trivially determined by geometric data alone: \begin{align}
\frac{rb^\prime - (d+1)b}{L^2a}=\Xi &= \frac{2(d+1)ra^2b^\prime - (d+1)(d+2)a^2b-r^2ba^{\prime2}-r^2\left(aa^\prime b^\prime+aa^{\prime\prime}b+a^2b^{\prime\prime}\right)}{2L^2a^3} \notag \\
&\;\;\;\;+\frac{\kappa^2 r^4}{2e^2L^4a^2}\left(p^{\prime2}Z + \tilde{p}^{\prime2}\tilde{Z}\right) - \frac{r^2\Phi^{\prime2}b}{aL^2}- \frac{V}{2L^2}  \label{veq}
\end{align}   The left equality sign holds on shell;  the right equality is the definition of $\Xi$.   It is easy to see that\begin{equation}
R_{rr} + \frac{R_{tt}}{b^2} = -\frac{da^\prime}{ra} = 2\Phi^{\prime 2},
\end{equation}or \begin{equation}
a(r) = \exp\left[-\frac{2}{d}\int\limits_0^r\mathrm{d}r^\prime\; r^\prime \Phi(r^\prime)^{\prime 2}\right] \label{aint}
\end{equation}where we have imposed the boundary condition $a(0)=1$.   To determine $b$, we use that: \begin{align}
\frac{R_{tt}}{b} &= \frac{2(d+1)ba^2 -draba^\prime -r^2 ba^{\prime 2}-(d+2)ra^2b^\prime + r^2aa^\prime b^\prime + r^2 aba^{\prime\prime} + r^2a^2 b^{\prime\prime}}{2r^2a^2} \notag \\
&= \frac{\kappa^2}{e^2}  \frac{Zp^{\prime 2}+\tilde{Z}\tilde{p}^{\prime 2}}{a} \frac{r^2}{L^2} - \frac{\Xi L^2 a}{r^2} = \frac{\kappa^2}{e^2}  \frac{Zp^{\prime 2}+\tilde{Z}\tilde{p}^{\prime 2}}{a} \frac{r^2}{L^2} - \frac{rb^\prime - (d+1)b}{r^2}.
\end{align}This leads to \begin{equation}
 a b^{\prime\prime} + ba^{\prime\prime} + a^\prime b^\prime - \frac{a^{\prime 2} b}{a} - \frac{d}{r}\left(ab^\prime + ba^\prime \right) = r^d a \left(\frac{(ab)^\prime}{ar^d}\right)^\prime  = 2\frac{\kappa^2r^2}{e^2L^2} \left(\tilde{Z}\tilde{p}^{\prime2}  + Zp^{\prime2}\right).   \label{eq39}
\end{equation}
The remaining two equations of motion are Maxwell's equations, which are best written in terms of Gauss' Law \cite{lsb}: \begin{equation}
-e^2\mathcal{Q} =    \frac{L^{d-2} Z p^\prime}{r^{d-2}a}, \;\; -e^2\tilde{\mathcal{Q}} =    \frac{L^{d-2} \tilde{Z} \tilde{p}^\prime}{r^{d-2}a}.   \label{zeq}
\end{equation} We also have the dilaton equation: \begin{equation}
 4\left(\frac{b}{r^d} \Phi^\prime\right)^\prime = \frac{a}{r^{d+2}} \frac{\partial V(\Phi)}{\partial \Phi} - \frac{\kappa^2 }{e^2 L^2r^{d-2} a} \frac{\partial Z}{\partial \Phi} p^{\prime2} -\frac{\kappa^2 }{e^2 L^2r^{d-2} a} \frac{\partial \tilde{Z}}{\partial \Phi} \tilde{p}^{\prime2}  \label{dilatoneq}
\end{equation}Using Maxwell's equation, we can simplify Eq. (\ref{eq39}): \begin{equation}
\left(\frac{(ab)^\prime}{ar^d}\right)^\prime = -2\frac{\kappa^2\mathcal{Q}}{L^d}p^\prime -2\frac{\kappa^2\tilde{\mathcal{Q}}}{L^d}\tilde{p}^\prime.  \label{peq}
\end{equation}

\subsection{Solutions at All Temperatures}
Let us now describe an efficient strategy for constructing interpolating geometries that (at $T=0$) interpolate from AdS as $r\rightarrow 0$, to a hyperscaling-violating geometry as $r\rightarrow\infty$.   This strategy will also allow us to construct the geometries at any temperature $T$.   

To manifestly ensure that our potentials are independent of $\mathcal{Q}$ and the couplings $e$, $L$ and $\kappa$, let us rescale \begin{equation}
r  \equiv \hat{r}\mathcal{\hat{Q}}^{-1/d},
\end{equation}(recall $\hat{\mathcal{Q}}$ was defined in Eq. (\ref{qhat})) and \begin{equation}
p \equiv \hat{p}  \mathcal{\hat{Q}}^{1/d} \frac{eL}{\kappa}
\end{equation}We use an identical rescaling for $\tilde{p}$.   It is easy to see from the EMD equations above that this change manifestly scales out $\mathcal{Q}$, $e$, $L$ and $\kappa$ from all of the equations,\footnote{I.e., in the equations before this subsection, we can ``set $\mathcal{Q}=e=L=\kappa=1$" and find a solution.} so any solution we find for $V(\Phi)$, $Z(\Phi)$ and $\tilde{Z}(\Phi)$ will not depend on the choice of $\mathcal{Q}$. For simplicity we temporarily remove the hats in the manipulations below. 

This manipulation reduces the problem to dimensionless ordinary differential nonlinear equations.  However, we wish to find a \emph{family} of solutions to these equations, parameterized by finite temperature $T$.  For this, we exploit the following trick.   There is a dimensionless ratio that persists in the Einstein-matter system:  $\tilde{\mathcal{Q}}/\mathcal{Q}$.  Suppose that we rescale $\tilde{\mathcal{Q}} \rightarrow q\tilde{\mathcal{Q}}$.   If we also rescale $\tilde{p} \rightarrow q\tilde{p}$, and split $b$ into \begin{equation}
b = b_0  + q^2 b_2,
\end{equation}keep $a$, $\Phi$, $V$ and $Z$ independent of $q$, and find a solution to the following $q$-independent equations: \begin{subequations}\label{eq29}\begin{align}
-2p^\prime &= \left(\frac{(ab_0)^\prime}{ar^d}\right)^\prime, \label{eq29a} \\
\frac{a^\prime}{a} &= -\frac{2r}{d}\Phi^{\prime 2}, \\
V &= \frac{d}{a}\left(rb_0^\prime - (d+1)b_0\right) + \frac{r^{d+2}p^\prime}{a}, \\
 \left(\frac{ab_2}{r^{2d}}\right)^\prime &= c\frac{a}{r^d}  \label{eq29d} \\
\left(\frac{b_2}{r^{d+1}}\right)^\prime &= -\frac{\tilde{p}^\prime}{d}, \\
Z &= -\frac{ar^{d-2}}{p^\prime}, \\
\tilde{Z} &= -\frac{ar^{d-2}}{\tilde{p}^\prime}, 
\end{align}\end{subequations}then for \emph{any choice of} $q$, we have constructed a solution to the nonlinear EMD system.   Our family of solutions obtained by tuning $q$ will correspond to the finite temperature family of solutions.    For simplicity, we choose to set $q=1$ when $\mathcal{Q}=\tilde{\mathcal{Q}}$.    We should note that the quantities $p$ and $\tilde{p}$ require constant shifts as $q$ is increased, so that $p(r=r_{\mathrm{h}}) = \tilde{p}(r=r_{\mathrm{h}})=0$.  However, the electric fields $p^\prime$ and $\tilde{p}^\prime$ are independent of $r_{\mathrm{h}}$.    Here and below, the functions $b_2$ and $\tilde{p}^\prime$ are defined to be $q$-independent, though one must remember that the matter content is $q$-dependent.

The constant $c$ can be chosen by choosing that the asymptotics of $b_2$ are \begin{equation}
b_2 = -r^{d+1} + \cdots, \;\;\; (r\rightarrow\infty).
\end{equation}This enforces \begin{equation}
c=d-1+\frac{d(z-1)-\theta}{d-\theta}.
\end{equation}

We have a set of 7 non-linear equations to solve, and we have 9 functions to find -- this means that there are 2 functions that we may input by hand into these equations.   We have found it easiest to choose $b_2$ and $p^\prime$ to obtain physically sensible results.  We also choose to set $q=0$ at $T=0$.   This means that, in the low temperature geometries, the new charged sector of the theory decouples, and also greatly relaxes the list of asymptotic conditions that must be satisfied to find a good solution.

We impose a rather strict list of boundary conditions and asymptotics on our solutions to these equations -- most of them are required for consistency of the solution, and the rest we have chosen simply for convenience in later analysis.  In the UV ($r\rightarrow 0$) these are $Z(0)=\tilde{Z}(0)=1$, $V(0) = -d(d+1)$,  $\Phi(r\rightarrow 0) \sim r^{d-1}$, $a(0)=b_0(0) = 1$,  $-b_2(r\rightarrow 0) \sim r^{d+1}$, and $p^\prime(r\rightarrow 0) = \tilde{p}^\prime(r\rightarrow 0) = -r^{d-2}$.\footnote{In the case $d=2$, these asymptotics mean that the subleading power of $a$ in the IR is $r^2$, and grows parametrically faster than the next-to-leading order term in $b(r)$, $\sim r^3$.   This makes extraction of thermodynamic quantities more tedious, but since the dilaton mode we are sourcing is normalizable the thermodynamics of these black holes is physically sensible (see e.g. \cite{vegh}).   In the cases $d>3$ (which we do not consider), this choice implies that the stress tensor is traceless.}    In the IR ($r\rightarrow\infty$) we find, along with $b_2 \approx -r^{d+1}$: \begin{subequations}\begin{align}
a(r) &\approx  a_\infty r^{-(d(z-1)-\theta)/(d-\theta)}, \\
b_0(r) &\approx  b_\infty r^{-(d(z-1)+\theta)/(d-\theta)}.
\end{align}\end{subequations}  $a_\infty$ and $b_\infty$ are undetermined.   This implies that \begin{equation}
\Phi(r) \approx \Phi_\infty \log \frac{r}{r_0}
\end{equation}
with $r_0$ related to the subleading behavior of $a(r)$ and \begin{equation}
\Phi_\infty \equiv\sqrt{\frac{d(z-1)-\theta}{2(1-\theta/d)}} .
\end{equation}   The gauge field is given by (at $T=0$) \begin{equation}
p \approx p_\infty r^{-d-dz/(d-\theta)}
\end{equation}with \begin{equation}
p_\infty = \frac{d(z-1)}{d-\theta}b_\infty.
\end{equation}For the potentials, we find \begin{subequations}\begin{align}
V(\Phi) &\approx -V_0 \mathrm{e}^{-\beta\Phi}, \\
Z(\Phi) &\approx Z_0 \mathrm{e}^{\alpha\Phi},
\end{align}\end{subequations}where \begin{subequations}\begin{align}
\alpha &= (d(d-\theta)+\theta) \sqrt{\frac{8}{d(d-\theta)(dz-d-\theta)}}, \\
\beta &= \theta  \sqrt{\frac{8}{d(d-\theta)(dz-d-\theta)}},
\end{align}\end{subequations}and \begin{subequations}\begin{align}
V_0 &= \frac{d^2}{a_\infty} \left(1+\frac{z}{d-\theta}\right)\left(1+\frac{z-1}{d-\theta}\right)  b_\infty  r_0^{-\beta\Phi_\infty}, \\
Z_0 &= \frac{a_\infty}{dp_\infty (1+z/(d-\theta))}r_0^{\alpha\Phi_\infty}.
\end{align}\end{subequations}
Finally, the requirement that $a(0)=1$ fixes the leading order behavior of $b_2$ to be \begin{equation}
b_2(r\rightarrow 0) \approx -\frac{c}{d-1}r^{d+1} + \frac{r^{2d}}{d(d-1)} + \cdots
\end{equation}as can be seen from analysis of Eq. (\ref{eq29d}).  The behavior of $\tilde{p}$ and $\tilde{Z}$ is not related to the original EMD system and is specific to the choice of $b_2(r)$.

We chose the UV asymptotics that we did for $\Phi$ because it makes it easy to construct solutions where $V(\Phi)$, $Z(\Phi)$, $\tilde{Z}(\Phi)$, and even $B(\Phi)$ numerically resemble analytic functions of $\Phi$ for all $\Phi>0$, and at large $\Phi$ that these functions will grow no faster (or slower) than $\sim \exp[\# \Phi]$.    From a bottom up perspective, this is not really a requirement, but it does make these solutions seem less strange.

In addition, we must have physical requirements, such as $\Phi^\prime$ being real.  This implies that $a$ is a decreasing function, which is also sufficient to obey the null energy condition \cite{edgar2011}.   The requirement that $Z$ and $\tilde{Z}$ are finite at all finite $\Phi$ implies that $p^\prime$ and $\tilde{p}^\prime$ must always be negative;  this in turn implies that $b_2 r^{-1-d}$ is an increasing function, and imposes constraints on $b_0$ and $a$ as well.    Finally, we would like for the location of the horizon -- the value of $r$, $r_{\mathrm{h}}$, such that $b(r_{\mathrm{h}})=0$, to be a strictly decreasing function of the Hawking temperature $T$ -- also the temperature of the boundary theory.   As the Hawking temperature is given by \begin{equation}
4\pi \hat T = b^\prime(r_{\mathrm{h}})
\end{equation} with \begin{equation}
\hat T \equiv T\hat{\mathcal{Q}}^{-1/d}
\end{equation}this provides us with a final constraint on our system.

When numerically constructing our series of functions, we may use these asymptotics to choose solutions in the UV and IR.   These also serve as checks that our methods are convergent, though the numerical methods required to solve these equations are so elementary we have never seen any issues.   Although we have used numerical methods to help solve our differential equations, more efficient ``gauges" for gravity may be fruitful in finding exact analytic solutions. 

This method cannot allow us to construct solutions saturating the bound \begin{equation}
z=1+\frac{\theta}{d}.
\end{equation}This is because the above constraints would require $b_2 = -r^{d+1}$, which requires $a=1$, and therefore $\Phi=0$.   We also note that in this case, there are often logarithmic corrections to the IR asymptotics of the fields described above \cite{edgar2011}.   One of the only known examples of a non-Fermi liquid, the Ising-nematic quantum critical point in $d=2$, has $\theta=1$ and $z=3/2$, saturating this bound; these exponents appear to be robust to loop corrections \cite{maxsubir}.   A holographic construction of a finite temperature family of these geometries, with a UV completion, is thus of interest, but requires a different technique.   

Let us make one final comment before proceeding.    Although the presence of the second gauge field does not mix with other sectors of the theory for any computations presented in this paper -- all of which are static ($\omega=0$) properties of these geometries -- this will not be true in general.    One might therefore ask whether there exists a geometry, purely within EMD theory, which is ``similar" to ours.   A simple check to perform is to compute\footnote{The tensor equation here is coordinate-dependent;  taking the trace one obtains a gauge-invariant result.   This does not qualitatively change the story.} \begin{subequations}\begin{align}
\Delta_{\mu\nu} &= \frac{1}{\kappa^2}\left(R_{\mu\nu} - \frac{R}{2}g_{\mu\nu}\right) -  T^{\mathrm{EMD}}_{\mu\nu}, \\
\Delta_\Phi &=  4\left(\frac{b}{r^d} \Phi^\prime\right)^\prime - \left(\frac{a}{r^{d+2}} \frac{\partial V(\Phi)}{\partial \Phi} - \frac{1}{r^{d-2} a} \frac{\partial Z}{\partial \Phi} p^{\prime2}\right)
\end{align}\end{subequations}with $T^{\mathrm{EMD}}_{\mu\nu}$ the stress tensor of the original EMD sector.   We typically find for the cases we have studied that $\Delta_{\mu\nu}$ is parametrically smaller than either of the terms on the right hand side in the UV and IR, and smaller by a factor $\sim 50$ in the transition region.   $\Delta_\Phi$ is parametrically small compared to either term in square brackets in the IR, but may become large compared to terms on the right hand side deep in the UV.  However, in this regime, the dilaton and gauge field do not back react strongly on the geometry, which is at leading order AdS-Schwarzchild.  Minor modifications of the $\Phi\rightarrow 0$ behavior of $V(\Phi)$ to ensure that both dilaton modes are normalizable will ensure that whatever dilaton modes are sourced do not strongly alter the background geometry.   So up to these minor corrections,  we have good approximate solutions to the EMD system without any extra gauge field!   

\subsection{Thermodynamics}
The field theories dual to black holes constructed by this technique have remarkably simple thermodynamics in many cases.    In particular, let us suppose (as we will choose in our $d=3$ example below) that in the UV ($r\rightarrow 0$): \begin{subequations}\begin{align}
a(r) &\approx 1 - \hat a r^{d+1}, \\
b_0(r) &\approx 1 - \hat b_0 r^{d+1}, \\
b_2(r) &\approx -\hat b_2 r^{d+1}.
\end{align}\end{subequations}This makes computing $\langle T^{\mu\nu}\rangle$ remarkably simple, and we find using the standard prescription \cite{deharo} the energy density \begin{equation}
\epsilon = \frac{d(\hat b_0 + \hat b_2) + (d+2)\hat a}{d+1}
\end{equation} and the pressure \begin{equation}
P = \frac{\hat b_0 + \hat b_2 - \hat a}{d+1}
\end{equation}When $\hat a \ne 0$, the stress tensor is no longer traceless.    This is generically equivalent to a non-vanishing dilaton background, so is consistent with expectations.    The chemical potential for the charge of interest is \begin{equation}
\mu = \mu_0 - p(r_{\mathrm{h}})
\end{equation}with $p(r)$ the integral of $p^\prime(r)$ with $p(\infty)\equiv 0$;  note $\mu_0=p(0)$ is the chemical potential at $T=0$.

In our construction, $\hat a$ and $\hat b_0$ are $T$-independent, and $\hat b_2$ changes in a simple way:   \begin{equation}
\hat b_2 = \hat b_{20} q^2,
\end{equation}with $\hat b_{20}$ again a $q$-indepdendent constant, and $q$ our tuning parameter from before which generates exact solutions.   $q$ can be converted into the horizon radius $r_{\mathrm{h}}$, and thus temperature $T$.   Assuming as before that $q=0$ at $T=0$, \begin{equation}
q = \sqrt{\frac{b_0(r_{\mathrm{h}})}{|b_2(r_{\mathrm{h}})|}}.
\end{equation}Also, note that if $\mu_0$ is the (non-vanishing) chemical potential for the conserved charge at $T=0$, that \begin{equation}
\hat b_0 = \frac{2 \mu_0}{d+1}.
\end{equation}

Let us now use our asymptotics for the geometries that interpolate from AdS to a hyperscaling violating geometry.   In the AdS regime, $b_0(r_{\mathrm{h}}) \approx 1$ and $r_{\mathrm{h}} \approx (d+1)/4\pi T$, as for the AdS black brane.   We find $q^2 \sim r_{\mathrm{h}}^{-d-1} \sim  T^{d+1}$, and that $\hat b_2$ dominates both $\epsilon$ and $P$.   This gives us Eq. (\ref{thermoeq1}).   In the deep IR, we instead have that $q^2 \sim r_{\mathrm{h}}^{-d-dz/(d-\theta)}$.   In fact, \begin{equation}
q^2 \approx b_\infty ^{-(d-\theta)/z} \left(\frac{4\pi T}{d+dz/(d-\theta)}\right)^{1+(d-\theta)/z},
\end{equation}which gives us the rest of Eq. (\ref{thermoeq1}).    

\subsection{Examples}
\begin{figure}
\centering
\includegraphics{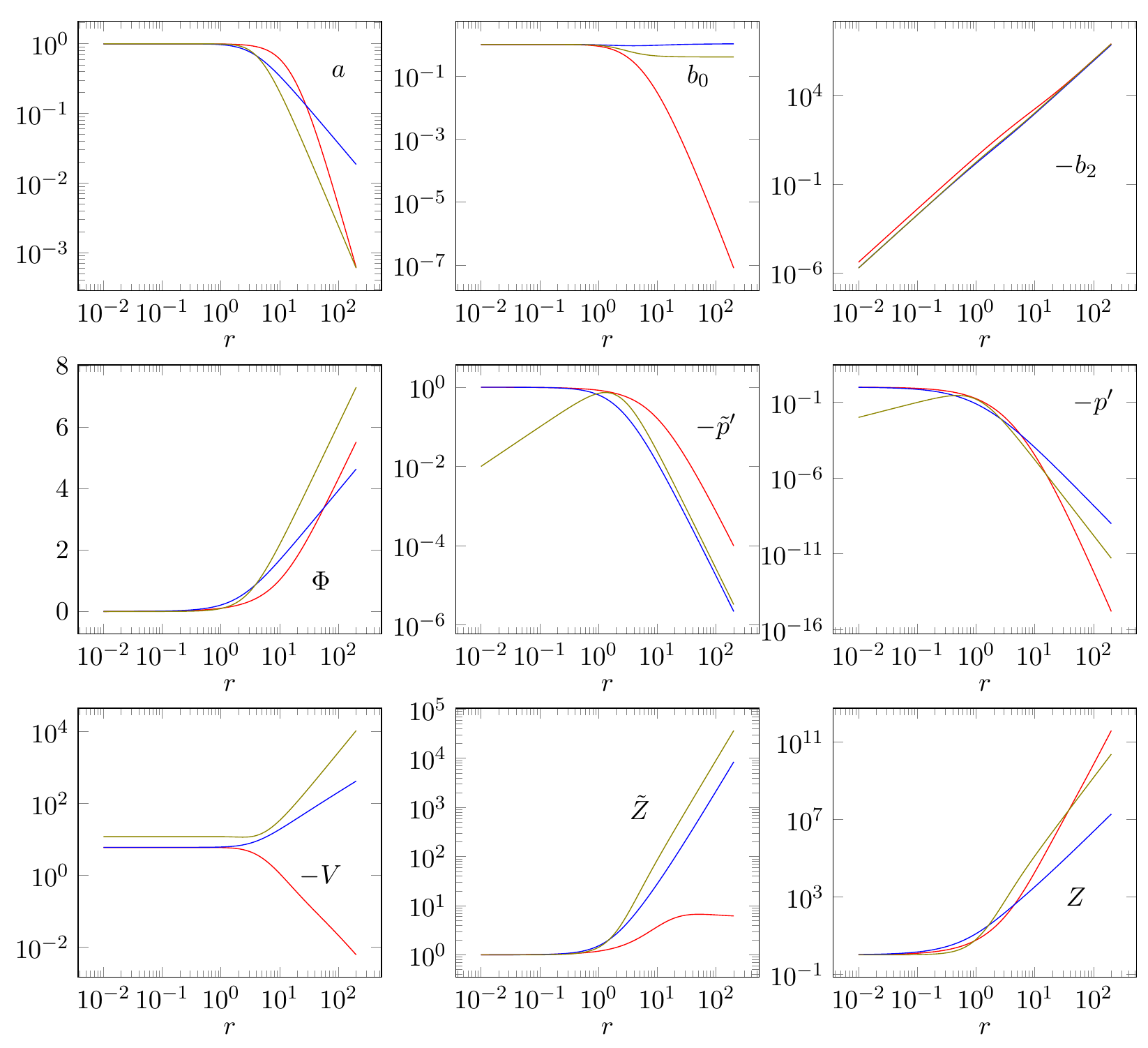}
\caption{The 9 functions of our construction corresponding to various matter fields.   We are taking the ``bare"  (no scaling with $q$) functions for $b_2$, $\tilde{p}^\prime$, and $\tilde{Z}$.  Red lines correspond to the geometry with $d=2$, $z=3$, $\theta=1$;  blue to $d=2$, $z=2$, $\theta=-2$;  olive to $d=3$, $z=\infty$, $\eta=3$.}
\label{matterfig}
\end{figure}

\begin{figure}
\centering
\includegraphics{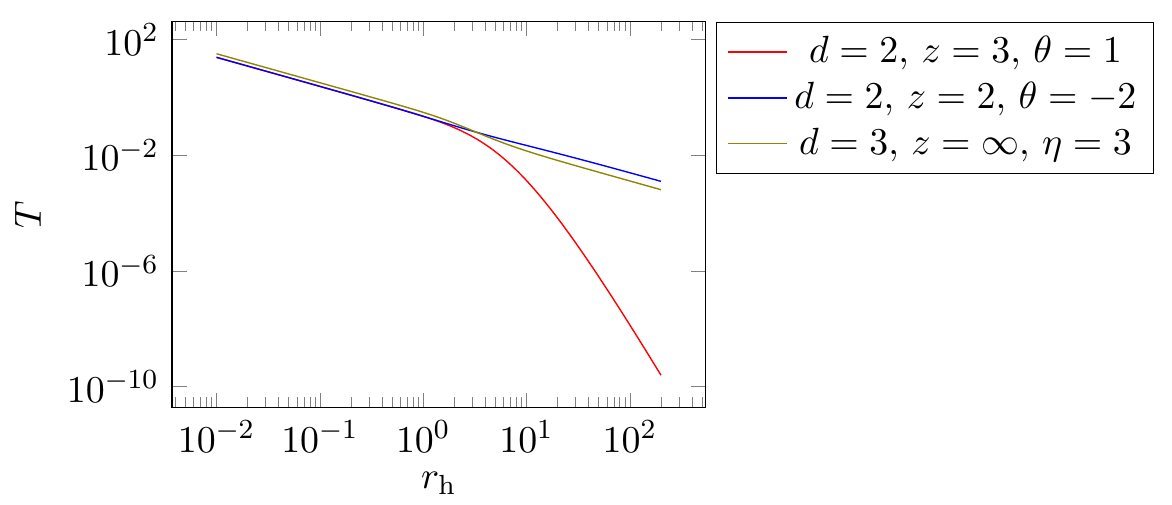}
\caption{The horizon radius as a function of temperature $T$ for the three geometries.}
\label{tfig}
\end{figure}

Here we present some numerical plots, and analytical details, about the specific three geometries we presented results for in the main text.    We begin with some plots;  analytical results are in following paragraphs.  Plots of all matter content in the background geometry are contained in Figure \ref{matterfig}.  A plot of $r_{\mathrm{h}}$ vs. $T$  is found in Figure \ref{tfig}.   

Importantly, we see that $T(r_{\mathrm{h}})$ is a monotonically decreasing function in all three numerical examples described below.   If on the other hand $T(r_{\mathrm{h}})$ has local extrema, the competition between the various phases described by different solutions may lead to discontinuous thermal phase transitions.   This is likely pathological and so we wish to avoid this effect.

For the $d=2$, $z=3$, $\theta=1$ geometry, we find that the constant $c=3$.   The choices \begin{equation}
b_2(r) = -r^3 \frac{r^2+12r+288}{r^2+12r+72}
\end{equation}and \begin{equation}
p^\prime(r) = -(1+ur)^{-9},\;\;\; u\approx 0.216
\end{equation}lead to a consistent solution with $\mu_0 \approx 0.578$.   The constant $u$ is numerically tuned so that $b_0(0)=1$ for this geometry; an analogous statement holds for subsequent geometries.   With the dilaton $\Phi \sim r$ in the UV,  and these choices for $b_2$ and $p^\prime$, we find that as $\Phi\rightarrow 0$, $Z,\tilde{Z} \approx 1+\#\Phi$ and $V\approx -6-\# \Phi^3$.   Without the dilaton-couplings $Z$,  the subleading coefficient of $V$ would be fixed by the UV asymptotics of $\Phi$ by the standard AdS/CFT dictionary, but this is not required once we include Maxwell fields.

For the $d=2$, $z=2$, $\theta=-2$ geometry, we find $c=1$.   The choices \begin{equation}
b_2(r) = -r^3 \frac{9r^2 +20r+80}{9r^2+20r+40}
\end{equation}and \begin{equation}
p^\prime(r) = -(1+ur)^{-4},\;\;\; u\approx 0.891
\end{equation}lead to a consistent solution with $\mu_0 \approx 0.374$.

For the $d=3$, $z=\infty$, $\theta=-3$ geometry, we find $c=1$.    The choices \begin{equation}
b_2(r) = -r^4 \frac{r^2+12}{r^2+6}
\end{equation}and \begin{equation}
p^\prime(r) = -r\left(1+u^2r^2\right)^{-3},\;\;\; u\approx 0.923
\end{equation}lead to a consistent solution with $\mu_0 \approx 0.293$.   We also find at at small $\Phi$, $Z$ and $\tilde{Z}$ are linear functions and $V$ is quadratic.  It is also helpful to know $r_{\mathrm{HV}}$ to compute IR operator dimensions for such a geometry;  we found $r_{\mathrm{HV}} \approx 4.9$.

Given our choices of $b_2(r)$, we solve for $a$, $\Phi$, $\tilde{p}^\prime$ and $\tilde{Z}$ in that order.\footnote{Not every step in this order (or in the subsequent one) must be done sequentially.}   With these results, and our choice for $p^\prime$, we find $p$, $b_0$, $Z$ and $V$.

We have checked that all of the power law (or logarithmic, in the case of $\Phi(r)$) growths of the matter content, evident in the IR in Figure \ref{matterfig}, is consistent with the general asymptotic analysis presented earlier.

\subsection{Coupling Neutral Scalars}
Since we have numerically constructed an interpolating geometry, we also need to numerically construct the interpolating function $B(\Phi)$ in Eq. (\ref{spsi}).    This may seem rather trivial at first glance, since we know the asymptotics of $B(\Phi)$:  see Eq. (\ref{Bphieq}).    It turns out, unfortunately, that generic choices of $B(\Phi)$, even if they behave correctly asymptotically, can induce instabilities whose signature is the presence of a normalizable mode of $\psi$ respecting the boundary conditions at the horizon.    As with superconductivity, these instabilities are associated with the interpolating region of the geometry.

To avoid this, we have numerically chosen an instability-free $B(\Phi)$ by choosing a function $\psi_{\mathrm{trial}}$ to solve the $k=0$, $T=0$ equations of motion.   The asymptotics on $\psi_{\mathrm{trial}}$ in this case are easily seen to be two distinct power laws, which up to Eq. (\ref{harriseq}) are arbitrary.   We simply take $\psi_{\mathrm{trial}}$ to be a sum of these two power laws with tunable relative coefficients\footnote{In the three examples we studied above, we chose the leading power law in the IR to be the one correctly associated to $\Delta_{\mathrm{IR}}$.} and compute \begin{equation}
B(r) = \frac{r^{d+2}}{a\psi_{\mathrm{trial}}} \left(\frac{b_0\psi_{\mathrm{trial}}^\prime}{r^d}\right)^\prime.
\end{equation}By construction, this choice of $B(\Phi)$ will have the correct asymptotics.

We have numerically found that these functions appear to track $V(\Phi)$ somewhat closely.   It may be the case that scalars with $B(\Phi) = \lambda V(\Phi)$, with fixed constant $\lambda$, may be ``better behaved" (for example, more natural from a top down construction).

\subsection{Large Chemical Potential}
We have found in our numerical examples above that $\mu \sim \mathcal{Q}^{1/d}$ with an $\mathcal{O}(1)$ coefficient.   Similarly, the transition temperature between AdS and the hyperscaling violating geometry, $T_{\mathrm{trans}}\sim \mathcal{Q}^{1/d}$; again the proportionality coefficient is $\mathcal{O}(1)$.   Thus, $T_{\mathrm{trans}}\sim \mu$.

From the viewpoint of condensed matter physics, this may not be desirable, as the scale $\mathcal{Q}^{1/d}$ may be quite large.   It is known \cite{sokol,georges,fitz1,fitz2,raghu,aass14}, how to obtain a transition temperature $T\ll \mu$ near a quantum critical point without holography.    Within holography, this is rather challenging.   The reason is as follows.   (Again we revert to dimensionless variables (with dimension set by appropriate powers of $\mathcal{Q}^{1/d}$) to describe the geometry.)  We wish to find a geometry where $a\approx b_0\approx 1$ with, say, $r\sim 1$, while $p(0)=\mu \gg 1$.   Using Eq. (\ref{peq}) we find \begin{equation}
p(r)\approx p(0) - \int\limits_0^r \mathrm{d}r^\prime \; Z(\Phi(r^\prime)) r^{\prime (d-2)}. 
\end{equation}If $Z(\Phi)$ smoothly varying, then since $\Phi\approx 0$, $Z\approx 1$, and in particular $p(r)\approx \mu$ for $r \lesssim 1$.   This is a contradiction with our assumptions that $a(r)\approx b(r)\approx 1$, however, by Eq. (\ref{eq29a}).    Relaxing the assumption that $Z(\Phi)$ is smoothly varying, it is formally possible to maintain $b(r)\approx 1$ while having $p(0)=\mu \gg 1$, and $p(1)\sim \mathcal{O}(1)$.   However this suggests that $Z(\Phi)$ is finely tuned to the value of $\mu$, which does not seem satisfactory.

Alternatively, one might imagine that there is an intermediate AdS regime, distinct from the true UV AdS.   The transition from this intermediate AdS to the IR hyperscaling violating geometry could then be at an energy scale very small compared to $\mathcal{Q}^{1/d}$.  The absorption of electromagnetic flux by a cloud of charged matter allows for AdS-to-AdS domain wall solutions \cite{gubsercloud1, gubsercloud2}; see also \cite{sera} for other AdS-to-AdS geometries.

\section{Computation of $\rho_{\mathrm{dc}}$ in Holography}\label{rhodcappendix}
\subsection{Transport Computation}
Here we outline the computation of $\rho_{\mathrm{dc}}$ in holography.   Here, we are referring to the resistivity associated with the bulk gauge field $A$.   We also demonstrate that, for this calculation, $\tilde{A}$ \emph{decouples} at leading order, and does not affect $\rho_{\mathrm{dc}}$.    This means that the result of \cite{Lucas} is still valid for our black holes despite the presence of a second gauge field.

We will follow, for the most part, the same logic as \cite{Lucas}, and so details that are skipped here may be found in \cite{Lucas}.    In order to compute the conductivity $\sigma_{\mathrm{dc}} = 1/\rho_{\mathrm{dc}}$, we employ the formula \begin{equation}
\sigma_{\mathrm{dc}} = \frac{L^{d-2}}{e^2}\lim_{\omega\rightarrow 0} \frac{-1}{\mathrm{i}\omega}\lim_{r\rightarrow 0} r^{2-d} \frac{\delta A_x^\prime(0)}{\delta A_x(0)}.  \label{defsigmadc}
\end{equation}where $\delta A_x(r)\mathrm{e}^{-\mathrm{i}\omega t}$ is a small perturbation about the equilibrium solution.    The perturbation $\delta A_x$ generically couples to all spin 1 perturbations in the problem.   These enclude $\delta \tilde{A}_x$, $\delta g_{tx} r^2/L^2 \equiv \delta \tilde{g}_{tx}$ (we set $\delta g_{rx}=0$ by a gauge choice), and a tower of spin 1 modes associated with the disordered scalar $\psi$.   Indeed, every single momentum mode of $\psi(\mathbf{k},r)$ may receive corrections -- we find it convenient to write out \begin{equation}
\psi_0(\mathbf{k},r) + \delta \psi(\mathbf{k},r) = \psi_0(\mathbf{k},r)\left(1+\delta P(\mathbf{k},r) \mathrm{e}^{-\mathrm{i}\omega t}\right)
\end{equation}and keep track of the perturbations $\delta P$ instead.    

If one linearizes the EMD system, in the $\omega\rightarrow 0$ limit one obtains a simple closed set of equations analogous to \cite{Lucas} \begin{subequations}\begin{align}
\frac{eL}{2\kappa ar^d} \delta\tilde{g}^\prime_{tx} &= \hat{\mathcal{Q}}\delta A_x + \hat{\tilde{\mathcal{Q}}}\delta \tilde{A}_x - L\kappa e \delta \mathcal{P}_x ,  \label{eq61a} \\
\delta\mathcal{P}_x^\prime &= - \frac{\delta \tilde{g}_{tx}}{dbr^d} \int \frac{\mathrm{d}^d\mathbf{k}}{(2\pi)^d} k^2 \psi_0(\mathbf{k},r)^2 + \mathcal{O}(\omega), \label{eq61b} \\
0 &= \left(\frac{eL \hat{\mathcal{Q}}}{\kappa}\delta\tilde{g}_{tx} - r^{2-d}bZ\delta A_x^\prime\right)^\prime + \mathcal{O}\left(\omega^2\right), \label{eq61c} \\
0 &= \left(\frac{eL \hat{\tilde{\mathcal{Q}}}}{\kappa}\delta\tilde{g}_{tx} - r^{2-d}b\tilde{Z}\delta \tilde{A}_x^\prime\right)^\prime + \mathcal{O}\left(\omega^2\right)  \label{eq61d}
\end{align}\end{subequations} where we have defined \begin{equation}
\delta \mathcal{P}_x \equiv \frac{b}{\omega r^d} \int \frac{\mathrm{d}^d\mathbf{k}}{(2\pi)^d} k_x \psi_0(\mathbf{k},r)^2 \delta P(\omega,\mathbf{k},r)^\prime.
\end{equation}Eqs. (\ref{eq61c}) and (\ref{eq61d}) can be integrated as $\omega\rightarrow 0$: \begin{subequations}\begin{align}
C &= \frac{eL \hat{\mathcal{Q}}}{\kappa}\delta\tilde{g}_{tx} - r^{2-d}bZ\delta A_x^\prime, \label{eq62a} \\
\tilde{C} &= \frac{eL \hat{\tilde{\mathcal{Q}}}}{\kappa}\delta\tilde{g}_{tx} - r^{2-d}b\tilde{Z}\delta \tilde{A}_x^\prime,   \label{eq62b}
\end{align}\end{subequations}where $C$ and $\tilde{C}$ are $\omega$-dependent constants.

Near the horizon $r=r_{\mathrm{h}}$, where $b(r) \sim T(r_{\mathrm{h}}-r)$,  we find that our equations are only consistent if the following scalings hold: \begin{equation}
\delta A_x \sim \delta \tilde{A}_x \sim \delta \mathcal{P}_x  \sim (r_{\mathrm{h}}-r)^{-\mathrm{i}\omega/4\pi T}.
\end{equation}Using Eq. (\ref{eq61b}) in Eq. (\ref{eq62a}), evaluating at the horizon, and taking the $\omega\rightarrow 0$ limit, we find at leading order \begin{equation}
C = -\mathrm{i}\omega r_{\mathrm{h}}^{2-d}Z(r_{\mathrm{h}}) \delta A_x(r_{\mathrm{h}}) - \mathrm{i}\omega  \frac{eL\hat{\mathcal{Q}}}{\kappa}\left(\frac{1}{dr_{\mathrm{h}}^d}\int \frac{\mathrm{d}^d\mathbf{k}}{(2\pi)^d} k^2\psi_0(\mathbf{k},r_{\mathrm{h}})^2\right)^{-1} \delta \mathcal{P}_x(r_{\mathrm{h}}) \sim \omega.   \label{eq65}
\end{equation}and an analogous result that $\tilde{C} \sim \omega$. We take the $r$-derivative of Eq. (\ref{eq61a}), and using Eqs. (\ref{eq61b})-(\ref{eq61d}) find a linear equation for $\delta \tilde{g}_{tx}$ that is sourced by only $C$ and $\tilde{C}$ (multiplying $r$-dependent coefficients) -- this ensures that $\delta \tilde{g}_{tx} \sim \omega$.
Eqs. (\ref{eq62a}) and (\ref{eq62b}) thus imply $\delta A_x^\prime , \delta\tilde{A}_x^\prime \sim \omega$, which in turn implies that in the $\omega\rightarrow 0$ limit, $\delta A_x$ and $\delta \tilde{A}_x$ are approximately $r$-independent constants.  
   
Now, because we are computing the conductivity associated with the conserved current dual to $A$, and not $\tilde{A}$, we must set $\delta \tilde{A}_x(0)=0$ -- from the boundary theory perspective, there is no electric field for this conserved current.    This implies that $\delta \tilde{A}_x$ may be neglected in the above formulas relative to $\delta A_x$, as both fields are approximately constants.   This is the only additional step in the calculation that differs from that in \cite{Lucas}, up to carrying around the extra gauge field.   We may finish the calculation.   Eq. (\ref{eq61a}) reduces to a simple proportionality relation between $\delta A_x(r_{\mathrm{h}}) = C_0$, and $\delta \mathcal{P}_x(r_{\mathrm{h}})$.   Combining Eqs. (\ref{defsigmadc}) and (\ref{eq65}) we obtain our answer, Eq. (\ref{eqsigmadc}).   

Importantly, the second gauge field has decoupled, implying that we may compute some reasonable dc transport properties of our theories while employing gauge fields in the bulk to help support the interpolating geometry.   At finite frequency, or for other transport coefficients (such as resistivity, which is the matrix inverse of the conductivity), the two gauge fields will mix, complicating the story.    

\subsection{Scaling Theory}
Let us begin by estimating $\rho_{\mathrm{dc}}$ when $T\gg \hat{\mathcal{Q}}^{1/d}$.   In this case, the geometry is governed by the asymptotically AdS regime, and is approximately an AdS-Schwarzchild black hole.   In this case, we find $\psi(\mathbf{k},r_{\mathrm{h}})$ by demanding regularity near the horizon \cite{Lucas}.   For momenta $k\gg T$, the solution $\psi \approx \mathrm{e}^{-kr}$ once $r\gg 1/k$ --  these high momenta are exponentially suppressed, the integral over $k^2\psi^2$ will be convergent, and in particular dominated by $k\lesssim T$.   For these small momenta, where $\psi$ is not small near the horizon, near the horizon, the near-horizon solution to the equation of motion is approximately \begin{equation}
\psi \approx \mathcal{C}_{\mathrm{h}} \mathrm{I}_0 \left(\mathcal{A} \sqrt{1-\frac{r}{r_{\mathrm{h}}}}\right) L^{-d/2}  \label{psih}
\end{equation}with $\mathrm{I}_0$ the modified Bessel function, $\mathcal{A}\sim \mathcal{O}(1)$, and $\mathcal{C}_{\mathrm{h}}$ a constant to be determined.   The overall prefactor of $L^{-d/2}$ is chosen so that $\mathcal{C}_{\mathrm{h}}$ is dimensionless.\footnote{We note that by dimensional analysis we may write the bulk scalar $\psi = \tilde{\psi}L^{-d/2}$, where $\tilde{\psi}$ is dimensionless.   The asymptotics on $\tilde{\psi}$ are that as $r\rightarrow 0$, $\tilde{\psi} = \tilde h(x)r^{d+1-\Delta}$, where $h$ is proportional to the source $h(x)$ coupling to the operator dual to $\psi$.   Again, dimensional analysis implies that $\tilde h$ and the source $h$ are identical up to an O(1) constant, which we can set to unity by a normalization choice of our boundary operator.}    We now need to match this onto a solution for $r\ll r_{\mathrm{h}}$, in asymptotically AdS.   Let us begin by focusing on $\mathbf{k}=\mathbf{0}$.   In the asymptotically AdS regime, there are power law solutions:  $\psi \sim r^{d+1-\Delta}$ or $r^\Delta$.  We write the UV behavior of $\psi$ as \begin{equation}
\psi \approx \left[c_1 \left(\frac{r}{r_{\mathrm{h}}}\right)^{d+1-\Delta} + c_2 \left(\frac{r}{r_{\mathrm{h}}}\right)^\Delta\right]L^{-d/2}.
\end{equation}$c_{1,2}$ are dimensionless constants that we must match up to the near-horizon geometry by matching $\psi$ and $\partial_r\psi$ when $r\sim r_{\mathrm{h}}$.   Of course, at this scale $r/r_{\mathrm{h}}$ is O(1), and we conclude that $c_{1,2} \sim \mathcal{C}_{\mathrm{h}}$, up to O(1) $\Delta$-dependent constants.    When $r\ll r_{\mathrm{h}}$, we can neglect the $c_2$ term.   Using the standard holographic dictionary, we conclude that $\varepsilon \sim \mathcal{C}_{\mathrm{h}}T^{d+1-\Delta}$.   We then conclude by noting that for $k \lesssim T$, the above argument is also valid -- momentum-dependent corrections to the background solutions only alter scaling for $r\gtrsim 1/k$ in a pure AdS background -- for a finite temperature black brane, what this means is simply that the finite temperature effects dominate over finite momentum effects.    Now, we are able to compute the $\psi$-dependent contribution to $\rho_{\mathrm{dc}}$: \begin{equation}
  \mathbb{E} \int \frac{\mathrm{d}^d\mathbf{k}}{(2\pi)^d} k^2 L^d \psi(\mathbf{k},r_{\mathrm{h}})^2 \sim  \mathbb{E} \int\limits_{|\mathbf{k}|<T} \frac{\mathrm{d}^d\mathbf{k}}{(2\pi)^d} k^2 L^d \psi(\mathbf{k},r_{\mathrm{h}})^2 \sim \mathcal{C}^2_{\mathrm{h}} T^{d+2} = \mathcal{C}^{\prime 2} \varepsilon^2 T^{2\Delta-d}.
\end{equation}where $\mathcal{C}^\prime$ is some O(1) constant.   Thus we find\footnote{Technically one might be concerned because we have been sloppy about averaging over disorder: e.g., we have assumed that $\mathbb{E}[1/\psi^2] \approx 1/\mathbb{E}[\psi^2]$.  At the level of a scaling argument, this is not problematic.} \begin{equation}
\sigma_{\mathrm{dc}} = \frac{1}{\rho_{\mathrm{dc}}} = \left(\frac{d+1}{4\pi T}\right)^{2-d} \frac{L^{d-2}}{e^2} + \mathcal{Q}^2\left(\frac{d+1}{4\pi T}\right)^{d}\frac{T^{d-2\Delta}}{\mathcal{C}^{\prime 2} \varepsilon^2}.
\end{equation}This can be easily inverted to obtain Eq. (\ref{rhosketch}), and the cross-over as $T$ is lowered from $\infty$ from a pair-creation dominated regime (where the first term in $\sigma_{\mathrm{dc}}$ dominates) to a scattering dominated regime (where the second term dominates) occurs at $T\sim T_{\mathrm{pc}}$, whose scaling is given in Eq. (\ref{tpc}).  

When $T\ll \hat{\mathcal{Q}}^{1/d}$, we enter the hyperscaling-violating region of the geometry.  We first consider the case when $z<\infty$.   Near the horizon, Eq. (\ref{psih}) is still valid.  A very similar matching argument can be used to determine $\mathcal{C}_{\mathrm{h}}$.    For $r \gg \hat{\mathcal{Q}}^{-1/d}$, and $k \lesssim \hat{\mathcal{Q}}^{(z-1)/dz}T^{1/z}$, which turns out to be the relevant criterion for when $\psi(\mathbf{k})$ is not exponentially suppressed near the black hole horizon, we can write $\psi$ as a sum of two power laws.   Away from the horizon, the smaller power law dominates \cite{Lucas}: \begin{equation}
\psi\left(\mathbf{k}, \hat{\mathcal{Q}}^{-1/d}\ll r\ll r_{\mathrm{h}}\right) \approx \mathcal{C}_{\mathrm{h}} \left(\frac{r}{r_{\mathrm{h}}}\right)^{(d+z-\Delta_{\mathrm{IR}}-\theta/2)/(1-\theta/d)}.
\end{equation}We next match this solution onto a UV solution.   Again, for $r\ll \hat{\mathcal{Q}}^{-1/d}$ only the leading power law will contribute, and we find that the near-boundary $\psi$ is given by \begin{equation}
\psi(\mathbf{k}, r\approx 0) \sim \mathcal{C}_{\mathrm{h}} L^{-d/2} \left(\frac{1}{\hat{\mathcal{Q}}^{1/d}r_{\mathrm{h}}}\right)^{(d+z-\Delta_{\mathrm{IR}}-\theta/2)/(1-\theta/d)}  \left(\hat{\mathcal{Q}}^{1/d}r\right)^{d+1-\Delta_{\mathrm{UV}}}
\end{equation}which, along with $\hat r_{\mathrm{h}} \sim \hat{\mathcal{Q}}^{-1/d}(T\hat{\mathcal{Q}}^{-1/d})^{-(1-\theta/d)/z}$, fixes \begin{equation}
 \mathcal{C}_{\mathrm{h}} \sim \varepsilon (T\hat{\mathcal{Q}}^{-1/d})^{-(d+z-\Delta_{\mathrm{IR}}-\theta/2)/z} \hat{\mathcal{Q}}^{(\Delta_{\mathrm{UV}}-d-1)/d}.
\end{equation} An identical argument to above gives us the remainder of Eq. (\ref{rhosketch}).

When $z=\infty$, and $T\rightarrow 0$, a similar matching procedure to the above gives us that $\psi(\mathbf{k},r_{\mathrm{h}}) \sim \mathcal{C}_{\mathrm{h}}$ with \begin{equation}
\mathcal{C}_{\mathrm{h}} \sim \varepsilon \hat{\mathcal{Q}}^{(\Delta_{\mathrm{UV}}-d-1)/d} (T\hat{\mathcal{Q}}^{-1/d})^{-\nu_-(\mathbf{k})(1-\theta/d)/z} \approx \varepsilon \hat{\mathcal{Q}}^{(\Delta_{\mathrm{UV}}-d-1)/d} (T\hat{\mathcal{Q}}^{-1/d})^{-(d+z-\Delta_{\mathrm{IR}}-\theta/2)/z + \mathcal{B}k^2},
\end{equation}where $\mathcal{B}\sim\mathcal{O}(1)$, and $\mathcal{B}>0$.   The above equation is valid for any mode with $k\lesssim \hat{\mathcal{Q}}^{1/d}$, where there will be no exponential damping of the mode before it enters the hyperscaling violating portion of the geometry.   We thus estimate (dropping the scaling in $\hat{\mathcal{Q}}$ and $\varepsilon$): \begin{equation}
\int \mathrm{d}^d\mathbf{k}\; k^2\psi^2 \sim T^{-(d+z-\Delta_{\mathrm{IR}}-\theta/2)/z} \int\mathrm{d}k\; k^{d+1} \mathrm{e}^{-\mathcal{B}\log(1/r_{\mathrm{HV}}T)k^2}.
\end{equation}The integral over $k$ gives the logarithmic corrections which appear in Eq. (\ref{rholog}).

\section{Neutral Scalars when $z=\infty$}\label{neuapp}
We consider the case where \begin{equation}
z\rightarrow \infty, \;\;\; \theta = -\eta z,
\end{equation}with $\eta$ finite.    Eq. (\ref{psiwave}) becomes \begin{equation}
r^2\psi^{\prime\prime} - \frac{d-\eta}{\eta} r\psi^\prime - \left(B_0 + k^2 r_{\mathrm{HV}}^2 \right)\psi  =0,  \label{powerlawpsi}
\end{equation}where $r_{\mathrm{HV}}$ is the ``radial scale at which the hyperscaling violating portion of the geometry begins."  We \emph{define} $r_{\mathrm{HV}}$ by\footnote{$r_{\mathrm{HV}}$ plays a role analogous to $\mu^{-1}$ in the $\mathrm{AdS}_2\times\mathbb{R}^2$ geometry, where operators also have $k$-dependent dimensions in the IR \cite{faulkner}.} \begin{equation}
g_{rr} \approx \frac{L^2}{r^2} \left(\frac{r_{\mathrm{HV}}}{r}\right)^2,\;\;\; (r\rightarrow\infty).
\end{equation}Note that for arbitrary $\eta$, $g_{rr}$ has this asymptotic behavior in the IR.  Eq. (\ref{powerlawpsi}) admits power law solutions $\psi \sim r^{\nu_\pm}$ with \begin{equation}
\nu_\pm = \frac{d(1+\eta^{-1}) \pm \sqrt{d^2(1+\eta^{-1})^2 + 4\left(B_0+(kr_{\mathrm{HV}})^2\right) }}{2}.
\end{equation}We are most interested in the non-normalizable solution with exponent $\nu_-$.     We find that this is $k$-dependent, and given by \begin{align}
\nu_-(kr_{\mathrm{HV}}) &= \frac{d}{\eta}\left(1+\frac{\eta}{2}-\frac{\Delta_{\mathrm{IR}}}{z}\right) + \frac{\sqrt{d^2(1+\eta^{-1})^2 + 4B_0}-\sqrt{d^2(1+\eta^{-1})^2 + 4\left(B_0+(kr_{\mathrm{HV}})^2\right)}}{2} \notag \\
&\approx \frac{d}{\eta}\left(1+\frac{\eta}{2}-\frac{\Delta_{\mathrm{IR}}}{z}\right) - \frac{2(kr_{\mathrm{HV}})^2}{(2\Delta_{\mathrm{IR}}/z - 1)(d/\eta)},\;\;\;(k\rightarrow 0).
\end{align}

\end{appendix}

\end{document}